\documentclass[useAMS,usenatbib]{mn2e}

\usepackage{fancyhdr}
\usepackage{graphicx,latexsym}
\usepackage{color}
\usepackage{amsmath}
\usepackage[margins]{trackchanges}
\usepackage{multirow}
\usepackage{subcaption}
\usepackage{subfig}
\usepackage{ctable}
\usepackage{url}
\usepackage{array}
\usepackage{color,soul}
\addeditor{LdB}
\addeditor{NS}
\addeditor{BB}
\addeditor{MS}

\usepackage{times}

\newcolumntype{M}{>{\centering\arraybackslash}m{\dimexpr.35\linewidth-2\tabcolsep}}
\newcolumntype{N}{>{\centering\arraybackslash}m{\dimexpr.16\linewidth-2\tabcolsep}}
\newcolumntype{Z}{>{\centering\arraybackslash}m{\dimexpr.23\linewidth-2\tabcolsep}}
\newcolumntype{B}{>{\centering\arraybackslash}m{\dimexpr.20\linewidth-2\tabcolsep}}

\title[Machine Classification of Transient Images]{Machine Learning Classification of SDSS Transient Survey Images}
\author[L. du Buisson et al.]{L. du Buisson$^{1,2}$\thanks{E-mail:
lisedubuisson@gmail.com}, N.
Sivanandam$^{2}$\thanks{E-mail:
navin.sivanandam@gmail.com}, Bruce A. Bassett$^{1,2,3}$\thanks{E-mail:
bruce.a.bassett@gmail.com} and M. Smith$^{4,5}$\\
$^{1}$Department of Mathematics and Applied Mathematics, University of Cape Town, Cross Campus Rd, Rondebosch 7700, South Africa\\
$^{2}$African Institute for Mathematical Sciences, 6-–8 Melrose Rd, Muizenberg 7945, South Africa\\
$^{3}$South African Astronomical Observatory, Observatory Rd, Observatory 7925, South Africa\\
$^{4}$Department of Physics, University of the Western Cape, Cape Town, 7535, South Africa\\
$^{5}$School of Physics and Astronomy, University of Southampton, Southampton, SO17 1BJ, United Kingdom}
\begin{document}

\date{Accepted 2015 September 2. Received 2015 August 30; in original form 2014 July 26}

\volume{454}
\pagerange{2026--2038} \pubyear{2015}

\maketitle

\label{firstpage}

\begin{abstract}
We show that multiple machine learning algorithms can match human performance in classifying transient imaging data from the SDSS supernova survey into real objects and artefacts. This is a first step in any transient science pipeline and is currently still done by humans, but future surveys such as LSST will necessitate fully machine-enabled solutions. Using features trained from eigenimage analysis (PCA) of single-epoch {\it g}, {\it r}, and {\it i}-difference images we can reach a completeness (recall) of $96\%$, while only incorrectly classifying at most $18\%$ of artefacts as real objects, corresponding to a precision (purity) of $84\%$. In general random forests performed best, followed by the k-nearest neighbour and the SkyNet artificial neural net algorithms, compared to other methods such as naive Bayes and kernel SVM. Our results show that PCA-based machine learning can match human success levels and can naturally be extended by including multiple epochs of data, transient colours and host galaxy information which should allow for significant further improvements, especially at low signal to noise. 
\end{abstract}


\section{Introduction}

The quest to answer the deepest open questions about the cosmos has pushed astronomers and cosmologists to sample larger and larger volumes of the Universe. Current and next-generation of surveys, such as GAIA\footnote{http://sci.esa.int/gaia/}, the Dark Energy Survey (DES)\footnote{http://www.darkenergysurvey.org/}, LSST\footnote{http://www.lsst.org} and the SKA\footnote{http://https://www.skatelescope.org/} will usher in an era of exascale astronomy requiring new machine learning and statistical inference tools.

The LSST, for example, will image the night sky with such depth and frequency that upwards of a million transient alerts are expected every night with at least one million Type Ia supernova (SNIa) candidates detected over a decade of operations \citep{p7}. This will swamp existing follow-up capabilities, pushing us into the era of photometric transient identification trained on small spectroscopic subsets \citep{p15, p16, p17, p18, p9, p10, p11}. Such techniques will always lead to a small set of misidentifications and the danger is that the resulting contamination, unless dealt with in a sophisticated way, will lead to biased results. 

However, long before one reaches the final scientific analysis, the data deluge create challenges in the analysis pipeline. For example, difference images are created by subtracting a reference image from the most recent image of a given part of the sky. In the ideal case this will leave a pure noise image unless a real transient such as a supernova, asteroid or variable star exists in the image. In reality there are inevitable artefacts that occur because of instrumental effects: diffraction spikes, CCD saturation and bleeding, registration errors and the like. 

We thus need to disentangle the potential objects of interest from the artefacts.  Historically this sorting and classification into real objects and artefacts has been done by astronomers scanning the images as soon as possible after the images have been taken. In the case of the SDSS supernova survey, this typically led to hundreds or thousands of images being scanned each night; a tedious job. Recently it has been shown that this hand scanning can be done effectively by crowdsourcing \citep{p6}; the public correctly identified $93\%$ of the spectroscopically-confirmed supernovae.  

However, using humans to do this classification makes it very difficult to quantify the biases that arise from the subtly different algorithms and internal decision trees in each human scanner's brain. In addition, the effective decision tree changes with time depending on the mood and tiredness of the hand-scanner\footnote{This is a well-known problem that affects even the most experienced practitioners. For example, the verdicts of judges vary strongly depending on the time since the last break \citep{p5}.} which obviously cannot be characterised systematically. This human bias was partially dealt with in the SDSS supernova survey by injecting fake SNe into the pipeline yielding an average detection efficiency for each scanner, but it is clearly a fundamental limitation of human hand scanning which is even worse for crowd-sourced classifications. Apart from these biases hand scanning will not be an option for LSST due to the millions of images that will need to be scanned each night. 

Replacing humans with machine learning for this transient-artefact classification therefore represents an important frontier in achieving the goals of future transient surveys. Existing work includes the pioneering work of the SNfactory \citep{p19, c1} where features included the position, shape and FWHM of the main object, as well as distance to the nearest object in the reference image. Also of interest are the work done by the Palomar Transient Factory (PTF), where the focus falls on distinguishing between transients and variable stars \citep{p23} and the discovery of variability in time-domain imaging surveys \citep{p24} where classifiers output probabilistic statements about the degree to which newly observed sources are astrophysically relevant sources of variable brightness. Recently \citet{p30} have presented a new Random Forest implementation for artefact/transient classification in the DES SN pipeline. Their feature set consisted of 38 features, most of which were based on analogs from \citet{p23} and \citet{p24}. Some of their new features were among the most important features for their classification - including flux and PSF based parameters that were obtained from their implementation of SExtractor \citep{p31}.

In contrast, we use SDSS data and derive our features from principal component analysis of the Sloan {\it g}, {\it r} and {\it i} difference images. We compare a number of different potential machine-learning algorithms such as k-nearest neighbours, artificial neural network (SkyNet -- \citet{p1}), naive Bayes and Support Vector Machine (SVM) and show that it is possible to achieve human-levels of classification completeness/recall with limited degradation in purity. 

In section \ref{data} we describe the SDSS data used; the testing and performance measures in section \ref{proc} and the feature extraction and machine learning algorithms we employed are described in sections \ref{feature} and \ref{MLT}. Our results are discussed in section \ref{results}.

\section{Data}
\label{data}
Our data is drawn from the 2nd and 3rd years\footnote{We do not use the first year of data since it used a very different set of criteria for selecting candidates.} of the SDSS-II supernova survey \citep{p20, p21} which operated in drift scan mode for three months each year from 2005-2007, alternately observing each of the two strips making up the approximately 300 deg$^2$ equatorial Stripe 82, weather permitting. In practise the mean cadence on any patch of sky was about four nights. 

The transient detection algorithm consists of image subtraction of the search image from a historical image of the same region of the sky in the Sloan colour bands {\it g}, {\it r}, and {\it i}. The difference image is flagged as a candidate for human scanning if it passes certain threshold cuts. However, this process was imperfect and led to a large number of artefacts that constituted $70\%$ of the candidates or more.    

\begin{figure}
\centering

\begin{subfigure}{.5\textwidth}
  \centering
  \includegraphics[width=\linewidth]{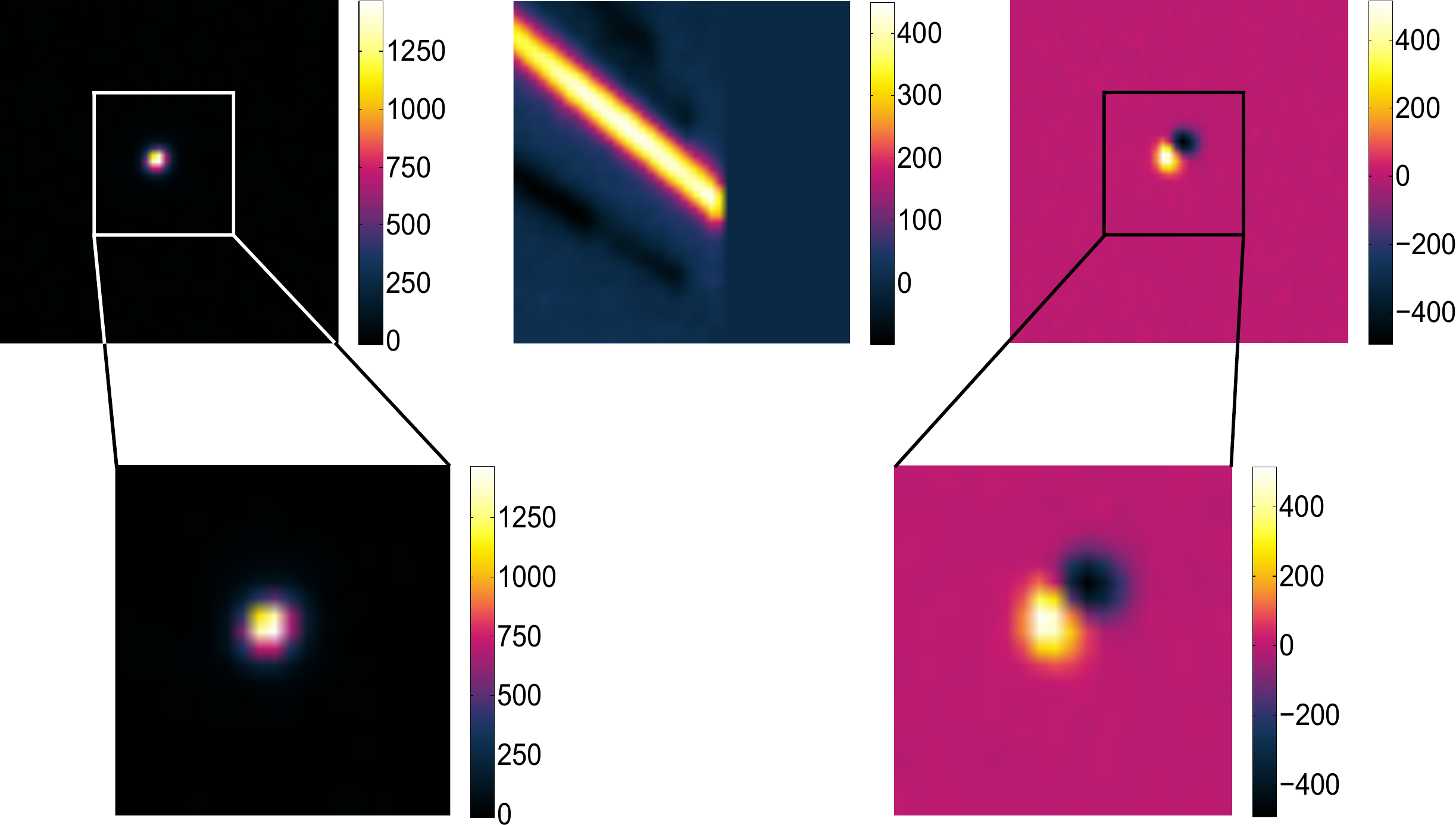}
  \caption{}
  \label{fig:lar_40}
\end{subfigure}%
\vspace{5mm}
\begin{subfigure}{.5\textwidth}
  \centering
  \includegraphics[width=\linewidth]{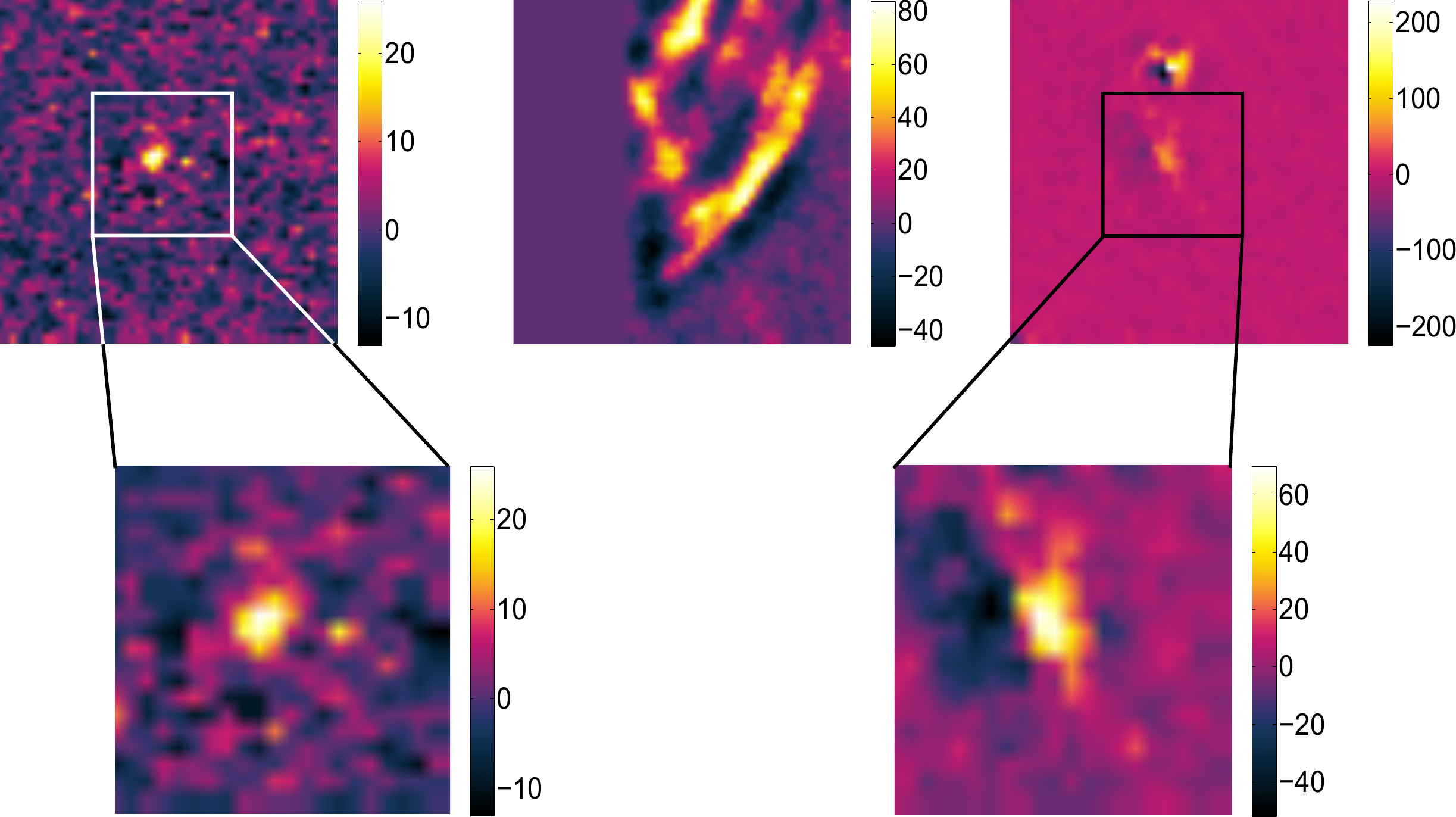}
  \caption{}
  \label{fig:sma_20}
\end{subfigure}
\caption{Example of the three different visual classes in the {\it r}-band used for our classification purposes; (a) shows high-quality images from the real (left), artefact (middle) and dipole/saturated (right) classes - all images here have an SNR above 40; (b) shows more representative images from the real (left), artefact (middle) and dipole/saturated (right) classes - all images here have an SNR below 20, representing roughly 80\% of the images in the dataset.}
\label{fig:data_all}
\end{figure}

A team of about 20 human hand scanners using both the search and difference images classified each candidate object into one of ten distinguishable classes: artefacts, dipoles, saturated stars, moving objects, variables, transients, SN Other, SN Bronze, SN Silver and SN Gold \citep{p20}. A description of each of these classes, as given by \citet{online1}, can be found in Table \ref{tab:class_cat}, and Table \ref{tab:hand_class} shows the visual appearance of each of these ten classes. For our classification purposes, we re-group these original classes into three new visual classes, based on the observation that many of the classes have very similar visual appearances. These three classes are: real objects, artefacts and dipoles/saturated, illustrated in Fig. \ref{fig:data_all}. Real objects have point-like residuals (convolved with the seeing from the atmosphere), artefacts have diffraction spike-like residuals while the dipoles/saturated class have \mbox{residuals} that are often close to point-like, but typically have negative flux in part of the difference image arising from registration errors or saturated CCD effects. Fig. \ref{fig:data_all} shows the three visual classes clearly: Fig. (\ref{fig:lar_40}) serves as an example of what high-quality images, with signal-to-noise ratios (SNRs) above 40, looks like, while Fig. (\ref{fig:sma_20}) shows images with SNR below 20, a threshold that represents roughly 80\% of the objects in our dataset. Table \ref{tab:class_cat} shows which of the original classes correspond to each of the three visual classes. This visual classification was used for carrying out single-class PCA for feature extraction, as is discussed in section \ref{feature}. We are ultimately concerned with whether we can correctly predict whether an object is real or not-real (artefact or dipole/saturated), however, and this is therefore the main classification that will be recognised by our classifiers. To see which classes correspond to real objects and which correspond to not-real objects, see Table \ref{tab:class_cat}.

The 2006 and 2007 seasons of the SDSS-II supernova survey introduced an automated software filter, {\it autoscanner}, to identify all objects detected in more than one epoch or bright objects detected for the first time and through statistical techniques, filter out non SN objects. This algorithm significantly reduced the number of moving objects and diffraction spikes, and eliminated a significant fraction of objects with long term variability by cross-matching with a veto catalogue. A detailed description of the {\it autoscanner} software is given in \citet{p25}.

As this algorithm was implemented after the 2005 season, our analysis considers only objects detected in the 2006 and 2007 seasons of the SDSS-II supernova survey. The dataset contains 27,480 objects, each consisting of three $51 \times 51$ pixel difference images (one for each of the {\it g}, {\it r} and {\it i} colour bands). Of this, 15,521 are real objects and 11,959 are not-real. For comparison, the 2005 season of SDSS-II detected 141,697 objects alone. The classifications used in this analysis are based on the classifications done by the SDSS hand scanners. About 2500 fake SNe were also inserted into the imaging pipeline in order to provide quality control and to characterise the selection function. Our dataset is the full set including fakes to allow us to compare with the human performance on the fakes.

\begin{table*}
\centering
 \caption{The different classifications for objects in our dataset. A description of each of the ten original classes used by the human hand scanners is given \citep{online1}, and their corresponding visual classification and main classification (real/not-real) is shown. See section \ref{data} for more detail.}
 \label{tab:class_cat}
 {\renewcommand{\arraystretch}{2}
 \begin{tabular}{@{}lm{5cm}cc}
  \hline
  Original Class & \multicolumn{1}{c}{Description} & Visual Class & Real/Not-Real \\
  \hline
  \hline
  {\bf Artefact} & Residuals caused by problems in the image (e.g. diffraction spikes). & Artefact & Not-Real \\
  
  {\bf Dipole} & Residuals with roughly equal amounts of positive and negative flux, caused by errors in image registration. & Dipole/Saturated & Not-Real \\
  
  {\bf Saturated Star} & Residuals of stars that saturate the CCD. & Dipole/Saturated & Not-Real \\
  
  {\bf Moving} & Anything showing signs of motion between cutouts in different passbands. & Real & Real \\
  
  {\bf Variable} & Objects showing a record of long-term variability. & Real & Real \\
  
  {\bf Transient} & Objects with no observation history, no apparent host galaxy in the search image, and no motion. & Real & Real \\
  
  {\bf SN Other} & Objects that are thought to have a good chance of being SNe, but that don't fit nicely into any of the above classes. & Real & Real \\
  
  {\bf SN Bronze} & Point-like residuals at the centre of their host galaxies - most of the objects in this class later turn out to be either quasars (QSOs), active galactic nuclei (AGN) or foreground variable stars, and not SNe. & Real & Real \\
  
  {\bf SN Silver} & Point-like residuals having no apparent host galaxy - SNe much more luminous than their host galaxies usually fall into this class. & Real & Real \\
  
  {\bf SN Gold} & Possible SNe identified as point-like residuals that are not at the exact centre of their host galaxies. & Real & Real \\
  \hline
 \end{tabular}}
\end{table*}

\section{Testing and Performance Measures}
\label{proc}
\subsection{Testing}
\label{testing}
Our basic testing protocol was to withhold 25\% of the data (hereafter referred to as the ``test set") for method comparison after our various learners have been trained, cross-validated (if necessary) and undergone preliminary testing using the other 75\% of the data. This 75/25 split was done prior to commencing building our various learning machines and the test set was sequestered until we were ready to perform our final testing.

Some of the classifiers have parameters that can be optimised using cross-validation. In general 30\% of the remaining 75\% of the data was kept back to perform the optimisation. This 30\% will be referred to as the ``validation set", while the remaining 70\% will be called the ``training set". We stress here that the sequestering of the true test data is essential for the unbiased comparison of our different classifiers. If one uses the same data to optimize an algorithm as to test there is a strong risk of ``training to the test set'', a phenomenon that will almost certainly lead to poor real world results.

\subsection{Performance measures}
\label{measures}
For both testing and for optimisation we have used four measures commonly applied to classification problems. These are the accuracy ($A$), the precision ($P$), the recall ($R$) and the $F_1$ score. They are defined in terms of the number of true/false positives/negatives ($t_p$, $t_n$, $f_p$ and $f_n$) as follows:  
\begin{align}
A &= \frac{t_p+t_n}{t_p+t_n+f_p+f_n}\\
P &= \frac{t_p}{t_p+f_p}\\ R &= \frac{t_p}{t_p+f_n}\\
F_1 & = \frac{2PR}{P+R}
\end{align}
For our problem, we define the positive classes to be those corresponding to real objects.

The choice of which of these measures should be used to measure success is rather dependent on the problem at hand. Accuracy can be a misleading measure in situations where the number of positive and negative cases are vastly different, e.g. a classifier that always predicts 1 will have an excellent accuracy on a problem where $99\%$ of cases are positive. For our data the classes are of similar sizes, so the accuracy is a reasonable initial measure. Somewhat better, however, is the $F_1$ score, which is expressed in terms of the precision (the fraction of reported real objects that are so) and the recall (the fraction of true objects that are found by the classifier). The $F_1$ score then is a measure that punishes false negatives and false positives equally, but weighted by their inverse fractional contribution to the full set to account for large class number hierarchies.

In real world applications, we are often more or less concerned with false positives than false negatives. For example, if the final number of items classified as objects is small enough to be easily verified by humans, the key will be to minimise $f_n$ and to maximise the recall. Conversely, if we are met with a data stream too large to be rechecked it may be more important to prevent contamination of data and so minimize $f_p$ and the precision.

We quote all of these statistics for completeness and also include the full confusion matrix, which contains the total numbers of true/false positives/negatives. An archetype is shown in Table \ref{confusion}.

\begin{table}
\centering
\caption{A schematic confusion matrix.}
\label{confusion}
{\renewcommand{\arraystretch}{1.5}
\begin{tabular}{c c c c}
 & & \multicolumn{2}{c}{{\bf Predicted Class}} \\
 & & Object & Not Object \\
 \hline
 \multirow{2}{20pt}{{\bf True Class}} & Object & $t_p$ & $f_n$ \\
 \cline{2-4}
  & Not Object & $f_p$ & $t_n$\\
 \hline
\end{tabular}}
\end{table}

Most classifiers can output the probability of an object being in a certain class and as a result we can study the trade-off between false positives and false negatives in a systematic way. The tool of choice for doing this is the Receiver Operating Characteristic (ROC) curve. This is a plot showing the performance of a binary classifier as its discrimination threshold is varied, and is created by plotting the True Positive Rate (TPR), or recall, versus the False Positive Rate (FPR) at various threshold values. The FPR, also known as the fall-out, can be calculated as shown in Eq. \ref{eq:fpr}.
\begin{align}
FPR = \frac{f_p}{f_p + t_n}
\label{eq:fpr}
\end{align}
Finally, the area under the ROC curve (``Area Under Curve" or AUC) is a statistic that is often used for model comparison in the machine learning community, and is equal to the probability that a classifier will correctly classify a randomly chosen instance \citep{p12}.

\section{Feature Extraction}
\label{feature}
Machine learning algorithms require features that represent the particular samples as their input. In the case of the transient classification problem our raw data are the pixel values for the images. In principle one could use the pixel values themselves as features, but this suffers from two problems. The first is a computational one; three $51\times51$ pixel images (for all three passbands) would give 7803 features. The second issue is that the images are highly compressible, so by including all the pixel data we run the risk of swamping the classifier with irrelevant features and suffering from the curse of dimensionality. 

To bypass these problems we used Principal Component Analysis (PCA) on the full training data set (multi-class PCA, section \ref{PCAoverall}) as well as on the individual visual classes (single-class PCA, section \ref{PCAclass}) and Linear Discriminant Analysis (LDA, section \ref{lda}) to extract those features that most faithfully represent the data.

\subsection{Principal Component Analysis}
\label{PCA}
Principal Component Analysis (PCA) is mainly used for the dimensionality reduction of data sets consisting of many correlated variables - data is orthogonally projected to a set of uncorrelated variables, referred to as Principal Components (PCs), ordered such that most of the variance present in the original data set is preserved in the first few PCs.

Defining a data set ${\bf X} = \{{\bf x}_1, {\bf x}_2, \ldots, {\bf x}_n\}$, where ${\bf x}_n$ is a $D$-dimensional observation with $n = 1, \ldots, N$, the aim of PCA is to project this data set onto a linear $M$-dimensional space, where $M < D$, such that most of the original data set's variance is preserved. To find these $M$ PCs, we first find the covariance matrix ${\bf S}$ of the data set, defined as usual as
\begin{align}
{\bf S} = \frac{1}{N} \sum_{n=1}^{N} ({\bf x}_n - \overline{{\bf x}}) ({\bf x}_n - \overline{{\bf x}})^\text{T},
\end{align}
where $\overline{{\bf x}}$ represents the mean of the data set, and then diagonalise {\bf S} to find the PCs (the eigenvectors) and eigenvalues of the covariance matrix. Selecting the $M$ largest PCs, a variable ${\bf x}_n$ can then be expressed in $M$-dimensional space by projecting it onto these PCs as
\begin{align}
{\bf a}_n = {\bf U}^{\text T}({\bf x}_n - \overline{{\bf x}})
\label{PCAeq}
\end{align}
with ${\bf a}_n$ the newly projected observation and ${\bf U}$ a $D \times M$ dimensional matrix with columns corresponding to the $M$ largest PCs.

In this paper we typically use the validation dataset to find the optimal value of $M$ for each method. We further use the Probabilistic PCA algorithm from the work of \citet{p22} which derives PCA from the perspective of density estimation and has a number of advantages for large datasets \citep{p26}.

\subsubsection{Creating Algorithm Inputs}
\label{PCAinputs}
To obtain the PC's (also referred to as ``eigenimages") of a set of $N$ objects $I_1, I_2, \cdots , I_N$, it is first necessary to represent each object $I_n$ by a corresponding vector, $\Gamma_n$. This is done by expressing each of the $g$, $i$ and $r$-band $N \times N$-pixel images as vectors and then concatenating them. If the $n$'th object's $g$-band image is denoted as
\begin{align}
I_{n-g} =
\begin{bmatrix}
  g_{1,1} & g_{1,2} & \cdots & g_{1,N} \\
  g_{2,1} & g_{2,2} & \cdots & g_{2,N} \\
  \vdots  & \vdots  & \ddots & \vdots  \\
  g_{N,1} & g_{N,2} & \cdots & g_{N,N}
\end{bmatrix},
\label{eq:I_n_g}
\end{align}
then its corresponding $g$-band vector is expressed as
\begin{align}
\Gamma_{n-g} =
\begin{bmatrix}
  g_{1,1} \\
  \vdots  \\
  g_{1,N} \\
  \vdots  \\
  g_{2,N} \\
  \vdots  \\
  g_{N,N}
\end{bmatrix}.
\label{eq:gamma_n_g}
\end{align}
The $i$ and $r$-band vectors are also obtained in this way. All three vectors are then concatenated to form the object's representative vector $\Gamma_n$, as
\begin{align}
\Gamma_n \quad = \quad
\begin{bmatrix}
  \Gamma_{n-g} \\
  \Gamma_{n-i} \\
  \Gamma_{n-r}
\end{bmatrix}
\quad = \quad
\begin{bmatrix}
  g_{1,1} \\
  \vdots  \\
  g_{N,N} \\
  i_{1,1} \\
  \vdots  \\
  i_{N,N} \\
  r_{1,1} \\
  \vdots  \\
  r_{N,N}
\end{bmatrix},
\label{eq:gamma_i}
\end{align}
and the data matrix ${\bf X}$ is then expressed as
\begin{align}
{\bf X} = 
\begin{bmatrix}
  \Gamma_1 & \Gamma_2 & \cdots & \Gamma_M
\end{bmatrix}
\label{eq:data}
\end{align}
with ${\bf X}^{\text T}$ being the input to the algorithm mentioned in section \ref{PCA}.

\subsubsection{Multi-class PCA}
\label{PCAoverall}
Multi-class PCA was done by carrying out PCA on the full training set (see section \ref{testing}) with all the classes mixed together. A small number of the resulting PCs were then kept for our feature extraction purposes. In order to derive the necessary features, all objects (from the test, validation and training sets) were first converted to vectors, as shown in section \ref{PCAinputs}, after which they were then expressed as a linear combination of the chosen PCs. The coefficients of these linear combinations were kept as the features of the objects (see Eq. \ref{PCAeq} for the calculation of these features for one object).

In Fig. (\ref{fig:pca_all}) one can see the first six principal components for the {\it r}-band images from the full training set. It can be seen that the last PCs in the figure are more noisy than the first ones. The optimal number of PCs used for feature extraction is dependent on the classifier at hand and is therefore a parameter that was optimised for every classifier by using various feature sets for training and validation. These various feature sets were derived by making use of either $0, \hspace{1mm} 5, \hspace{1mm} 10, \hspace{1mm} 25, \hspace{1mm} 50, \hspace{1mm} 100$ or $200$ PCs, respectively, so that the best set of features for each classifier could be determined. See section \ref{featuresets} for more details regarding our feature sets.

\subsubsection{Single-class PCA}
\label{PCAclass}
\begin{figure*}
\centering

\begin{subfigure}{.95\textwidth}
  \centering
  \includegraphics[width=\linewidth]{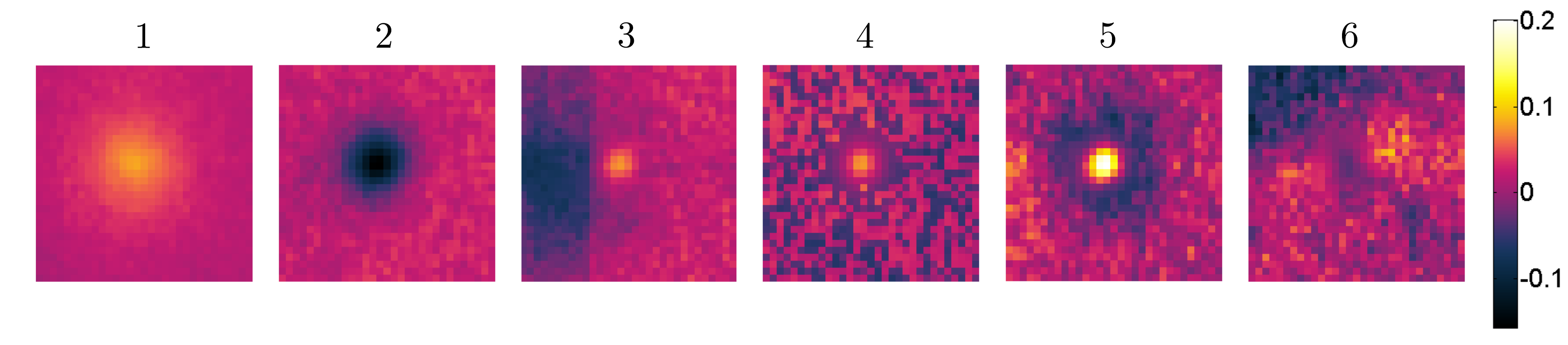}
  \caption{All classes.}
  \label{fig:pca_all}
\end{subfigure}%

\begin{subfigure}{.95\textwidth}
  \centering
  \includegraphics[width=\linewidth]{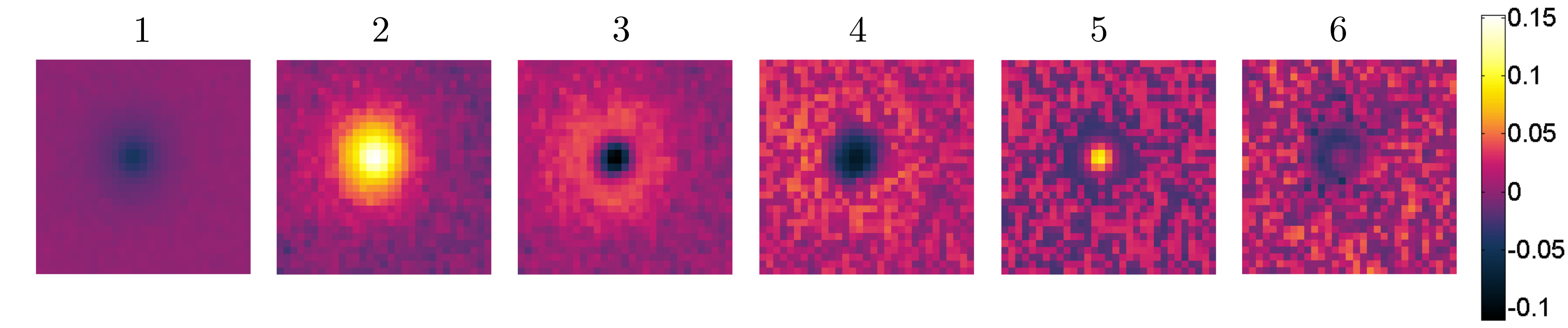}
  \caption{Real objects.}
  \label{fig:pca_real}
\end{subfigure}%

\begin{subfigure}{.95\textwidth}
  \centering
  \includegraphics[width=\linewidth]{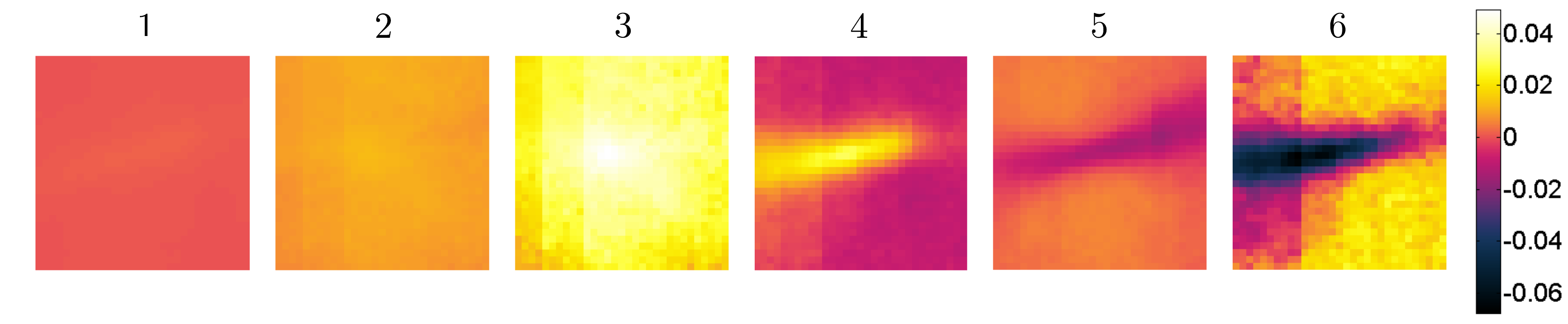}
  \caption{Artefacts.}
  \label{fig:pca_art}
\end{subfigure}%

\begin{subfigure}{.95\textwidth}
  \centering
  \includegraphics[width=\linewidth]{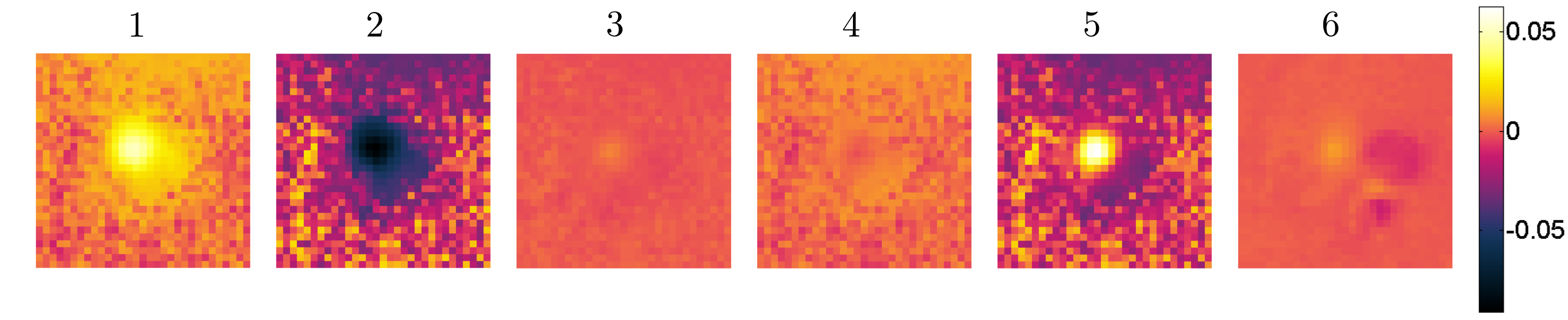}
  \caption{Dipoles/saturated.}
  \label{fig:pca_satdip}
\end{subfigure}
\caption{The first six PCA components in the {\it r}-band of (a) all the classes in the training data grouped together, (b) the real class, (c) the artefact class and (d) the dipole/saturated class.}
\label{fig:pca}
\end{figure*}

Another approach to feature extraction is to apply PCA independently to each of the three visual classes of objects (see Table \ref{tab:class_cat}) in the training data, yielding a unique set of eigenimages for each class (see Fig. \ref{fig:pca_real}, \ref{fig:pca_art} and \ref{fig:pca_satdip} for $r$-band PC's of the different visual classes). Features for an object $I_n$ are then obtained by first reconstructing it using the 15 largest PCs of each class in turn, and then calculating the error per pixel (taken as the Euclidean distance) between each reconstructed image and the original image, as
\begin{align}
\varepsilon_{n(class)} = \sqrt{\left( I_n - \widetilde{I}_{n(class)}\right) ^2 / m}
\label{eq:error}
\end{align}
where $m$ is the number of pixels in an object (including all three of its passbands) and $\widetilde{I}_{n(class)}$, with  $class \in \{ real, art, sat/dip \}$, is the reconstruction of the object $I_n$ using the PCs from the different visual classes in turn to yield three errors: $\varepsilon_{n(real)}$, $\varepsilon_{n(art)}$ and $\varepsilon_{n(sat/dip)}$. These three calculated errors are then used as features for the object.

\subsection{LDA}
\label{lda}
One potential issue with PCA for classification problems is that the direction of maximum variance may not align with the boundary between classes, the so-called decision boundary. There are several ways to alleviate this problem. The simplest is to include more PCA components in the features -- as $n_{\textrm{components}}$ increases so does the probability that the information required to separate the samples is contained.

Another possibility is to include components generated from Linear Discriminant Analysis (LDA) along with PCA features. LDA projects data in a direction that maximises the variance between classes while simultaneously minimising the variance within a class. LDA suffers from only being able to generate a number of components $\leq n_{\textrm{classes}}-1$, so can be insufficient as a class separator if used alone. We implemented the LDA algorithm from the work done by \citet{p22} who, in turn, based their design on that of \citet{b4}. We did LDA on the full training set; seeing as we classify objects into one of two possible classes, an object therefore only had one LDA component for use as a feature.

\subsection{Feature Sets}
\label{featuresets}
After obtaining features from PCA (namely the PC weights from multi-class PCA and the reconstruction errors from single-class PCA) and LDA our final step has been to normalise the data to give all the features a standard deviation of 0.5 and a mean of 0. This is done to improve the efficiency of the various routines that optimise the objective function for the different classifiers.

By varying the number of PC weights, omitting or including an LDA component, using either normalised or non-normalised features sets and deriving features using either uncropped images of $51 \times 51$ pixels or cropped images of $31 \times 31$ pixels, we created a total of 56 different feature sets. The various classifiers we have studied have been optimised by varying the feature sets as described above and carrying out intermediate testing. The final test results for a given classifier are reported using whichever feature set gave the best performance.

\section{Machine Learning Techniques}
\label{MLT}
\subsection{Minimum Error Classification}
\label{MEC}
Minimum Error Classification (MEC) is an extremely simple classifier, and is used here in order to form an idea of what the most basic of tools can achieve.

MEC takes only the three error-related features obtained from carrying out a single-class PCA (as described in section \ref{PCAclass}), and assigns an object to the class corresponding to the minimum reconstruction error (the calculation of which is described by Eq. \ref{eq:error}).

Intuitively, this simply reflects the logic that an image should, on average, belong to the class with the best PCA reconstruction (smallest error). We found that the best results were obtained when three visual classes were used for single-class PCA: real objects, artefacts and dipoles/saturated (as discussed in section \ref{data}). A sample is then classified as not being an object if the minimum error corresponds with either the artefact class or the dipole/saturated class. It is classified as a real object otherwise.

Because there is no training process involved, MEC was tested directly on the final test data. The results, shown in Table \ref{tab:summary}, was obtained by using the central $31 \times 31$ pixel subimages and a non-normalised feature set of reconstruction errors.

\subsection{Na{\"i}ve Bayes}
\label{NB}
Bayesian reasoning dictates that all quantities of interest are controlled by their probability distributions, and that by reasoning about these probabilities while taking observed data into account, we can make optimal decisions. The importance of this reasoning to machine learning lies in its ability to provide a quantitative approach to weighing the evidence for alternative hypotheses \citep{b1}.

The Na{\"i}ve Bayes (NB) classifier \citep{b1} decides on the most probable class $c_m$ via the computation:
\begin{align}
c_m &= \underset{c_j \in C}{\operatorname{argmax}} \hspace{1mm} P(c_j|a_1, a_2, ..., a_n)
\label{eq:3} 
\end{align} 
Here $C$ represents the finite set of class values (real and not-real objects) and $a_1, a_2, ..., a_n$ are the $n$ image feature values.
We then use Bayes theorem: 
\begin{align}
P(h|D) &= \frac{P(D|h) P(h)}{P(D)}
\label{eq:4}   
\end{align}
where $h$ is a given hypothesis and $D$ is the data, and assuming that $P(a_1, a_2, ..., a_n)$ is independent of $c_j$, we find  
\begin{align}
c_m &= \underset{c_j \in C}{\operatorname{argmax}} \hspace{1mm} P(a_1, a_2, ..., a_n|c_j) P(c_j)
\label{eq:5}
\end{align}  
Estimating the $P(c_j)$ priors is easily done by counting the frequency with which each class $c_j$ appears in the training data. To estimate the $P(a_1, a_2, ..., a_n|c_j)$ terms in the same fashion, however, requires a very large training set, and is therefore usually not feasible when there are many features due to the exponentially large volume that needs to be sampled.
 
NB classification rests on the simplifying assumption that, given the class value $c_j$, the feature values are conditionally independent, leading to the final expression: 
\begin{align}
c_m &= \underset{c_j \in C}{\operatorname{argmax}} \hspace{1mm} P(c_j) \prod_{i} P(a_i|c_j)
\label{eq:6}
\end{align}
Now, instead of having to estimate various $P(a_1, a_2, ..., a_n|c_j)$ terms as before, we need only estimate $P(a_i|c_j)$, a much more feasible process. Physically this corresponds to projecting the full posterior onto each of the features sequentially. Rather than assume a specific probability distribution for the estimation of $P(a_i|c_j)$, we binned the features and used the sample distribution of the training data to estimate $P(a_i|c_j)$ in each bin and for each class. Features were binned such that the number of samples per bin averaged four. It was found that varying the uniform bin sizes had little impact on the performance of the Na{\"i}ve Bayes algorithm - in terms of accuracy we checked that four samples per bin was optimal in a search up to 50 samples per bin.

The major limiting issue for NB lies in its assumption of the conditional independence of feature values - something that is rarely true for real life problems and often leads to the degradation of the performance of the classifier.

Applying this method to the various training and validation data sets revealed that using the central $31 \times 31$ pixel subimages with a non-normalised feature set comprising of 50 PCA weights and three reconstruction error values give the best results. The results obtained when applying NB to the corresponding test data can be seen in Table \ref{tab:summary}.

\subsection{K-Nearest Neighbours}
\label{knn}
K-Nearest Neighbours (KNN) is an exceptionally simple classifier. It finds the K-nearest training objects in feature space to a test point, averages their classes (with a uniform or distance weight) and classifies the test item accordingly (see the book by \citet{b4}).

The classifier has the advantages of being simple, inherently non-linear and free of intrinsic parameters. Performance can be poor in high-dimensional spaces however, but this was not a problem in our case. Initial training and validation were carried out with each of our 56 feature sets. Of the variations, the best result seemed to come from using the central $31\times31$ pixel subimages with 10 nearest neighbours, 53 features (50 PCA components and 3 reconstruction errors) and normalisation of the data. In this case the results for the training and validation sets were similar: accuracy $93\%$, precision $92\%$, recall $96\%$ and $F1$-score $94\%$. The results when this model was applied to the test data are given in Table \ref{tab:summary}. The KNN algorithm here implemented was taken from work done by \citet{p22}.

\subsection{Support Vector Machine}
\label{svm}
\begin{figure}
\centering
\includegraphics[width=0.5\textwidth]{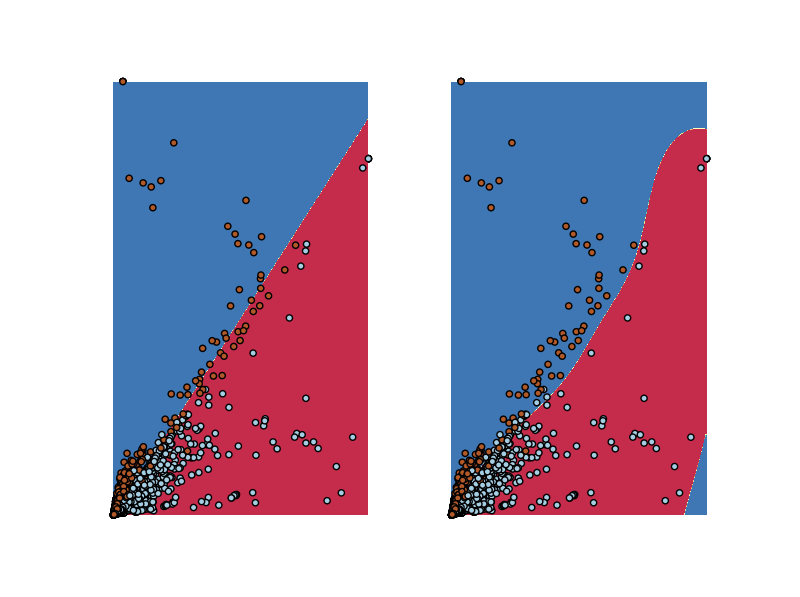}
\caption{Linear SVM vs Gaussian kernal SVM, for 2 PCA components.  The non-linear decision surface does a  better job of separating the classes which is even more true in the full feature space.}
\label{KernelPicture}
\end{figure}

The Support Vector Machine (SVM) is a maximum margin classifier -- it tries to find a decision boundary such that not only are the classes separated, but also that the separation is as large as possible. In the simplest implementation of an SVM, this decision boundary is a hyperplane in the feature space. The strength of the SVM lies, however, in the ease with which the so-called kernel trick can be applied to the data that maps the features into a higher dimensional space in which the classes are well-separated by a hyperplane. An illustrative example of the kernel trick mapping a linear boundary into a curved boundary is shown in Fig. \ref{KernelPicture}, allowing more successful separation of real from not-real objects. For detail on this classifier, see the book by \citet{b5}.

An SVM with a Gaussian (or radial basis function -- RBF) kernel function has two parameters that need to be optimized -- the width of the Gaussian (expressed in terms of $\gamma=1/2\sigma^2$, where $\sigma$ is the width) and the soft margin parameter ($C$). The latter sets the trade-off between a smooth decision surface (small $C$, less biased) and a better fit (large $C$, more accurate). These parameters are optimised using a grid search on quasi-exponentially increasing parameters and $n$-fold cross-validation -- the training set is split randomly into $n$ equal sets, the classifier is trained on $n-1$ of these, the error on the remaining set is calculated, the process is repeated for the $n$ choices of the test set, the average error is calculated and after repeating this procedure for every parameter value pair the $C$ and $\gamma$ values that minimise the cross-validation error are selected. This process is carried out for each of our 56 different feature sets. We chose to implement an SVM taken from the work done by \citet{p22}.

After cross-validation (and trying a variety of different feature sets) the optimal results on the training and validation data came from $C=1000$ and $\gamma=0.1$, using the central $31\times31$ pixel subimages with a feature set consisting of 100 normalised components from PCA, one LDA component and three reconstruction errors. The results when this model was applied to the corresponding test data can be found in \mbox{Table \ref{tab:summary}.}

\subsection{Artificial Neural Network}
\label{skynet}
\begin{figure}
\centering
\includegraphics[width=0.27\textwidth]{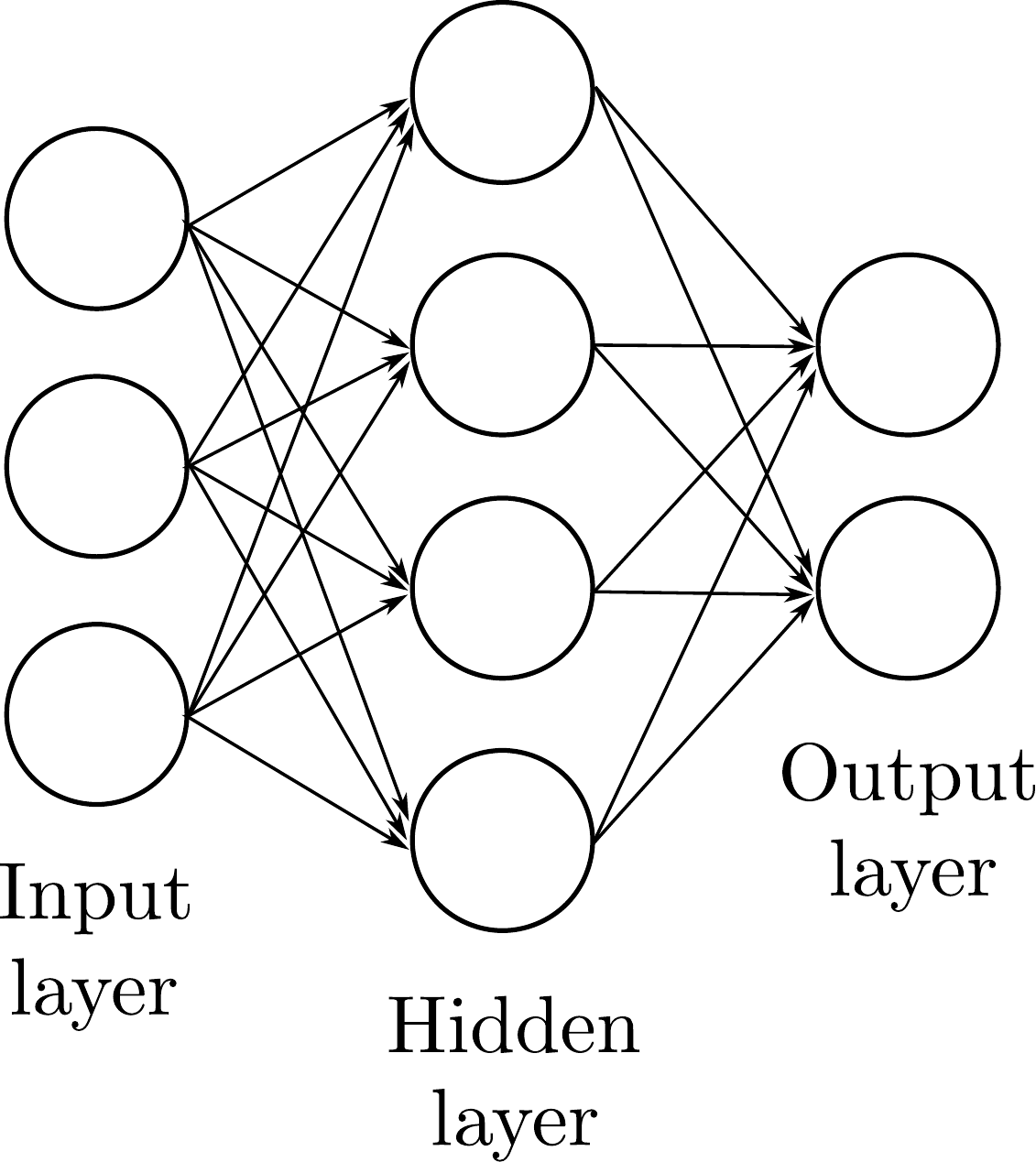}
\caption{A 3-layer feed-forward neural network consisting of a 3-node input layer, one 4-node hidden layer and a 2-node output layer.}
\label{fig:ANN}
\end{figure}

Loosely inspired by biological neural systems, artificial neural networks (ANNs) form a well-known class of machine learning algorithms. These networks consist of interconnected nodes, each of which receives and processes information before passing the result onwards to other nodes along weighted connections. Generally speaking, ANNs can be of arbitrary structure, but for many machine learning problems only feed-forward ANNs are required - networks that are directed from an input to an output layer of nodes and that have none, one, or many `hidden' layers in between, as shown in Fig. \ref{fig:ANN}. Using training data, the network `learns' a mapping between the inputs and the outputs, and is then able to predict the outputs for new input test data. \citep{p1}

We used SkyNet as our ANN as it is a well-known and robust neural network training tool developed by \citet{p1}. It can train large and deep feed-forward ANNs for use in a wide range of machine learning applications, e.g. regression, classification and dimensionality reduction, to name but a few. SkyNet has a few very advantageous features: It allows the training of an autoencoder, a feed-forward ANN where the inputs are mapped back to themselves - these networks provide an intuitive approach to non-linear dimensionality reduction. It also allows the training of recurrent neural networks (RNNs), a class of ANNs in which connections between nodes form a directed cycle, creating a network of nodes with feedback connections. Furthermore, SkyNet employs an optional pre-training method that obtains network weights and biases close to the global optimum of the network objective function, instead of starting the training process from a random initial state. SkyNet also makes use of convergence criteria that prevents it from overfitting to the training data.

\subsubsection{Number of hidden layers and nodes}
A universal approximation theorem \citep*{p2} states that feed-forward ANNs with at least three layers (one input, one hidden and one output layer), can approximate any continuous function to some given accuracy. This is the case for as long as the activation function is piecewise continuous, locally bounded, and not a polynomial - which is indeed the case for SkyNet.

Furthermore, empirical \citep{p3} and theoretical \citep{p4} considerations suggest that the optimal structure for the approximation of a continuous function is through the use of one hidden layer with $2N + 1$ nodes ($N$ being the number of input nodes). It was decided that these guidelines would be followed when setting up the ANNs for our classification problem for as long as it remained computationally feasible.

\subsubsection{Other network settings and classifier results}
For our classification purposes SkyNet was not configured to act as either an autoencoder or a RNN, and the pre-training method on offer was not used. Testing showed that enabling these options did not result in any significant changes in the classification performance for the problem at hand, and they were therefore deemed unnecessary. Furthermore, the inputs to the network were left unwhitened\footnote{Whitening here refers to the normalisation of the inputs to the network - an option when specifying settings in SkyNet. For more detail on how this is done for SkyNet, see \citet{p1}.}, and the average error-squared of the network outputs were used to determine convergence. For more details regarding these settings, see \citet{p1}.

We found that using the central 31 $\times$ 31 pixel subimages with a non-normalised feature set comprising 200 PCA weights and three reconstruction error values did the best for the training and validation phase. The results for the corresponding test set can be seen in Table \ref{tab:summary}.

For most of the smaller feature sets that were tested, a hidden layer with $2N + 1$ nodes were used, as previously discussed. For the larger feature sets (like the best-performing feature set mentioned above), however, that would result in a computationally unfeasibly large number of hidden nodes. In the case of the winning feature set (above), a hidden layer with 100 nodes was used, and 140 iterations were needed for the algorithm to converge.

\subsection{Random Forest}
\label{rf}
Random forests (RF), first introduced by \citet{p27}, are ensemble \mbox{classifiers} consisting of a collection of decision trees - they classify instances by combining the predictions of their trees together.

Each such tree is grown from a random bootstrapped sample (of equal size as the training set but selected with replacement) from the training set - this aggregation of bootstrapped samples, each one of which is used to grow a separate tree, is referred to as {\it bagging} \citep{p28}.

At each node of a tree, we need to decide how to split the data - this decision could be based on one or more feature values at a time. $K$-d trees (which were not used in our analysis), for instance, usually use only one feature value when deciding how to optimally split the data at each node. For random forests it is common practice to select $k \leq K$ feature values when determining the optimal split for nodes in their decision trees, where $K$ is the number of features in a dataset - the choice of $k$ is usually a parameter specified by the user. For the random forest implementation we used in this paper, additional randomness was added by selecting the optimal splitting point from a {\it random} sample of $k < K$ features at every node in a tree. The value of $k$ was chosen as $k = \sqrt{K}$, the suggested default value for the algorithm. As described by \citet{p29}, $k$ is related to the strength of a tree in the classifier (the stronger a tree, the lower its error rate) and the correlation between different trees (a forest with highly correlated trees will have a higher error rate).

To classify a new instance the classifier combines the individual predictions of all the trees in the forest, either by having them vote for the class that is the most popular or by averaging their probabilistic predictions (as was the case for our implementation). Adding more trees does not improve test performance beyond a certain point, implying that random forests are robust against overfitting.

We implemented the RF algorithm taken from the work of \citet{p22}. After training and validation using our variety of different feature sets and optimising for the number of decision trees (varied between 10 and 1000), we found that the best results were obtained for random forests consisting of 600 trees and above using the central $31 \times 31$ pixel subimages with a non-normalised feature set comprising 100 PCA components and three reconstruction errors. Applying a 600-tree random forest model to the corresponding test set yielded the results given in Table \ref{tab:summary}.

\section{Results}
\label{results}
\subsection{Best Performing Classifier}
For the supernova application we decided that accuracy and recall would serve as the most important performance metrics; see section \ref{measures}. Because the two main classes in our data (real objects and not-real objects) are of similar sizes, accuracy serves as a good general measure of performance, while recall was chosen because we are more concerned with missing true objects (false negatives) than we are with contaminating our set of predicted objects with false positives since these are easy for humans to weed out.

\begin{figure}
\centering
\includegraphics[width=0.5\textwidth]{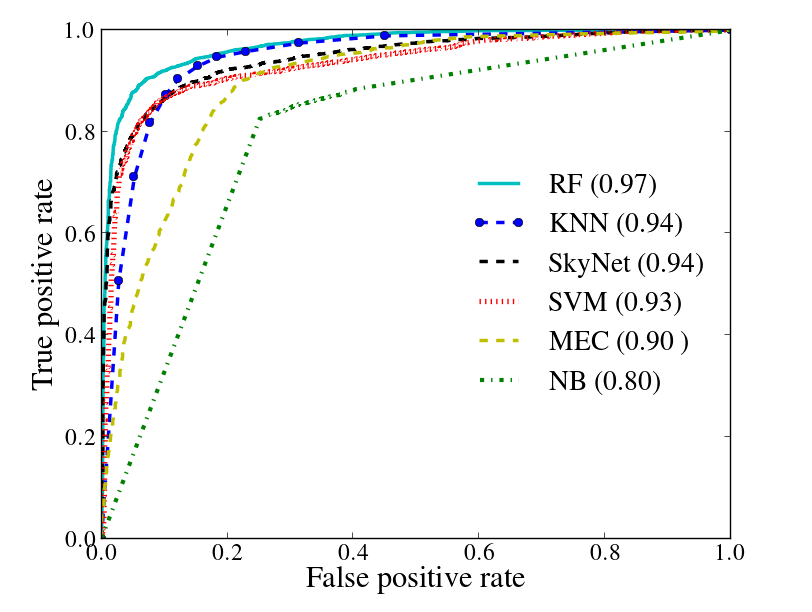}
\caption{The ROC curve and Area Under Curve (AUC) for each of the six classifiers. The AUC is indicated in brackets next to each classifier's name, and classifiers are listed in order from best-performing to worst-performing, based on the AUC statistic. It should be noted that changing the threshold value of RF, for example, can easily result in a recall (true positive rate) of $\sim96\%$ with only a slight penalty in the false positive rate, an encouraging sign.}
\label{fig:roc}
\end{figure}

The Receiver Operating Characteristic (ROC) curves for each of the six classifiers based on their performance on the final test set, is shown in Fig. \ref{fig:roc}, along with their Area Under the Curve (AUC) values. It is worth pointing out that by changing the threshold values of classifiers, the false and true positive rates can be adjusted according to the survey at hand - for large surveys like the LSST, for example, a small false positive rate will be required in order to efficiently deal with the number of false positives in the classified dataset.

Table \ref{tab:summary} gives a summary of the results of the different classifiers, and lists them in order from best-performing to worst-performing based on accuracy, recall and AUC value. It is interesting to note that random forests (RF) perform best in all metrics except recall, and further provides the best true positive rate at any value of the false positive rate (see Fig. \ref{fig:roc}). 
KNN, a simple nonparametric classifier, performed 2nd best in all metrics except recall, closely followed by SkyNet and SVM. Interestingly, the simplest algorithm, MEC, had the best recall, though it lagged significantly in its true positive rate at small values of the false positive rate; \mbox{Fig. \ref{fig:roc}.} 

How much can we trust the results in Table \ref{tab:summary}? One approach is to estimate the sample variance errors on the accuracies, recalls and precisions for each of the classifiers by using bootstrap resampling. The results produced through this procedure suggest sample variance errors of about $\pm 0.01$ at $95\%$ confidence, and hence our rankings of classifier performances are reasonably robust. However, the differences between the best classifiers (RF, KNN, SkyNet and SVM) are small and it is almost certainly the case that better optimisation of any of the algorithms could change the final ordering we obtained. However, from the point of view of replacing human image scanning, it is encouraging that multiple algorithms did very well. 

\begin{table*}
\centering
\renewcommand\multirowsetup{\centering}
 \caption{A summary of the performance results of the various classifiers, ordered from best-performing (top) to worst-performing (bottom). The best result for each performance metric is indicated in bold. The true labels are on the left side of the confusion matrices, and the predicted labels are at the top (see Table \ref{confusion}).}
 \label{symbols}
 {\renewcommand{\arraystretch}{1.5}
 \begin{tabular}{@{}ccccccp{0.3cm}ccc}
  \hline
  Machine Learning Technique & AUC & Accuracy & Recall & Precision & F1-score & & \multicolumn{3}{c}{ Confusion Matrix} \\
  \hline
  \hline
  \multirow{3}{80pt}{{\bf Random Forest \\(RF)} \\ \mbox{Section \ref{rf}}} & \multirow{3}{20pt}{{\bf 0.97}} & \multirow{3}{20pt}{{\bf 0.91}} & \multirow{3}{20pt}{0.91} & \multirow{3}{20pt}{{\bf 0.93}} & \multirow{3}{20pt}{{\bf 0.92}} & \multirow{3}{20pt}{} & & {\bf Object} & {\bf Not Object} \\
  \cline{8-10}
 & & & & & & & {\bf Object} & 3541 & 342\\
  \cline{9-10}
 & & & & & & & {\bf Not Object} & 259 & 2732 \\ 
  \hline
  \multirow{3}{80pt}{{\bf K-Nearest Neighbours \\(KNN)} \\ \mbox{Section \ref{knn}}} & \multirow{3}{20pt}{0.94} & \multirow{3}{20pt}{0.89} & \multirow{3}{20pt}{0.90} & \multirow{3}{20pt}{0.91} & \multirow{3}{20pt}{0.90} & \multirow{3}{20pt}{} & & {\bf Object} & {\bf Not Object} \\
  \cline{8-10}
 & & & & & & & {\bf Object} & 3506 & 377\\
  \cline{9-10}
 & & & & & & & {\bf Not Object} & 363 & 2628 \\ 
  \hline
  \multirow{3}{80pt}{{\bf SkyNet} \\ \mbox{Section \ref{skynet}}} & \multirow{3}{20pt}{0.94} & \multirow{3}{20pt}{0.88} & \multirow{3}{20pt}{0.89} & \multirow{3}{20pt}{0.90} & \multirow{3}{20pt}{0.89} & \multirow{3}{20pt}{} & & {\bf Object} & {\bf Not Object} \\
  \cline{8-10}
 & & & & & & & {\bf Object} & 3461 & 422\\
  \cline{9-10}
 & & & & & & & {\bf Not Object} & 399 & 2592 \\
  \hline
  \multirow{3}{90pt}{{\bf Support Vector Machine \\(SVM)} \\ \mbox{Section \ref{svm}}} & \multirow{3}{20pt}{0.93} & \multirow{3}{20pt}{0.86} & \multirow{3}{20pt}{0.90} & \multirow{3}{20pt}{0.85} & \multirow{3}{20pt}{0.87} & \multirow{3}{20pt}{} & & {\bf Object} & {\bf Not Object} \\
  \cline{8-10}
 & & & & & & & {\bf Object} & 3514 & 369\\
  \cline{9-10}
 & & & & & & & {\bf Not Object} & 605 & 2386\\
  \hline
  \multirow{3}{110pt}{{\bf Minimum Error Classification \\(MEC)} \\ \mbox{Section \ref{MEC}}} & \multirow{3}{20pt}{0.90} & \multirow{3}{20pt}{0.84} & \multirow{3}{20pt}{{\bf 0.92}} & \multirow{3}{20pt}{0.83} & \multirow{3}{20pt}{0.87} & \multirow{3}{20pt}{} & & {\bf Object} & {\bf Not Object} \\
  \cline{8-10}
 & & & & & & & {\bf Object} & 3559 & 324\\
  \cline{9-10}
 & & & & & & & {\bf Not Object} & 754 & 2237\\
  \hline
 \multirow{3}{80pt}{{\bf Na{\"i}ve Bayes \\(NB)} \\ \mbox{Section \ref{NB}}} & \multirow{3}{20pt}{0.80} & \multirow{3}{20pt}{0.77} & \multirow{3}{20pt}{0.86} & \multirow{3}{20pt}{0.77} & \multirow{3}{20pt}{0.81} & \multirow{3}{20pt}{} & & {\bf Object} & {\bf Not Object} \\
  \cline{8-10}
 & & & & & & & {\bf Object} & 3333 & 550 \\
  \cline{9-10}
 & & & & & & & {\bf Not Object} & 998 & 1993 \\
  \hline
 \end{tabular}}
 \label{tab:summary}
\end{table*}

\subsection{Inter-classifier Agreement}
To form an idea of the extent of inter-classifier agreement in terms of images correctly and incorrectly classified, we calculate the Cohen Kappa coefficient $\kappa$ \citep{p13}, a statistical measure of inter-classifier agreement argued to be more robust than a simple percent agreement calculation, as it also takes into account chance agreement. Cohen's Kappa measures the agreement between two classifiers that each classifies $N$ items into $C$ mutually exclusive classes, and is calculated as:
\begin{align}
\kappa = \frac{P(a) - P(b)}{1 - P(b)}
\label{eq:kappa}
\end{align}
where $P(a)$ is the observed relative  agreement and $P(b)$ is the hypothetical probability of chance agreement. If the classifiers agree perfectly, $\kappa = 1$, whereas if there is no agreement between classifiers except for that which would be expected by chance alone, then $\kappa = 0$. Statisticians differ slightly on the exact interpretation of different values of $\kappa$ (see the guidelines given by \citet{p14}, \citet{b2} and \citet{b3}, respectively), but for a rough idea as to what these interpretations are, see Table \ref{tab:interp}, the guidelines given by \citet{p14}.

\begin{table}
\centering
 \caption{Guidelines to the strength of agreement between two classifiers based on the value of the Cohen's Kappa coefficient $\kappa$ \citep{p14}.}
 \label{tab:interp}
 {\renewcommand{\arraystretch}{1.5}
 \begin{tabular}{@{}cc}
  \hline
  Kappa Statistic & Strength of Agreement \\
  \hline
  \hline
  $\kappa \leq 0.00$ & None\\
  $0.00 < \kappa \leq 0.20$  & Slight\\
  $0.20 < \kappa \leq 0.40$ & Fair\\
  $0.40 < \kappa \leq 0.60$ & Moderate\\
  $0.60 < \kappa \leq 0.80$ & Substantial\\
  $0.80 < \kappa < 1.00$ & Almost Perfect \\
  $\kappa = 1.00$ & Perfect \\
  \hline
 \end{tabular}}
\end{table}

\begin{figure}
\centering
\includegraphics[width=0.45\textwidth]{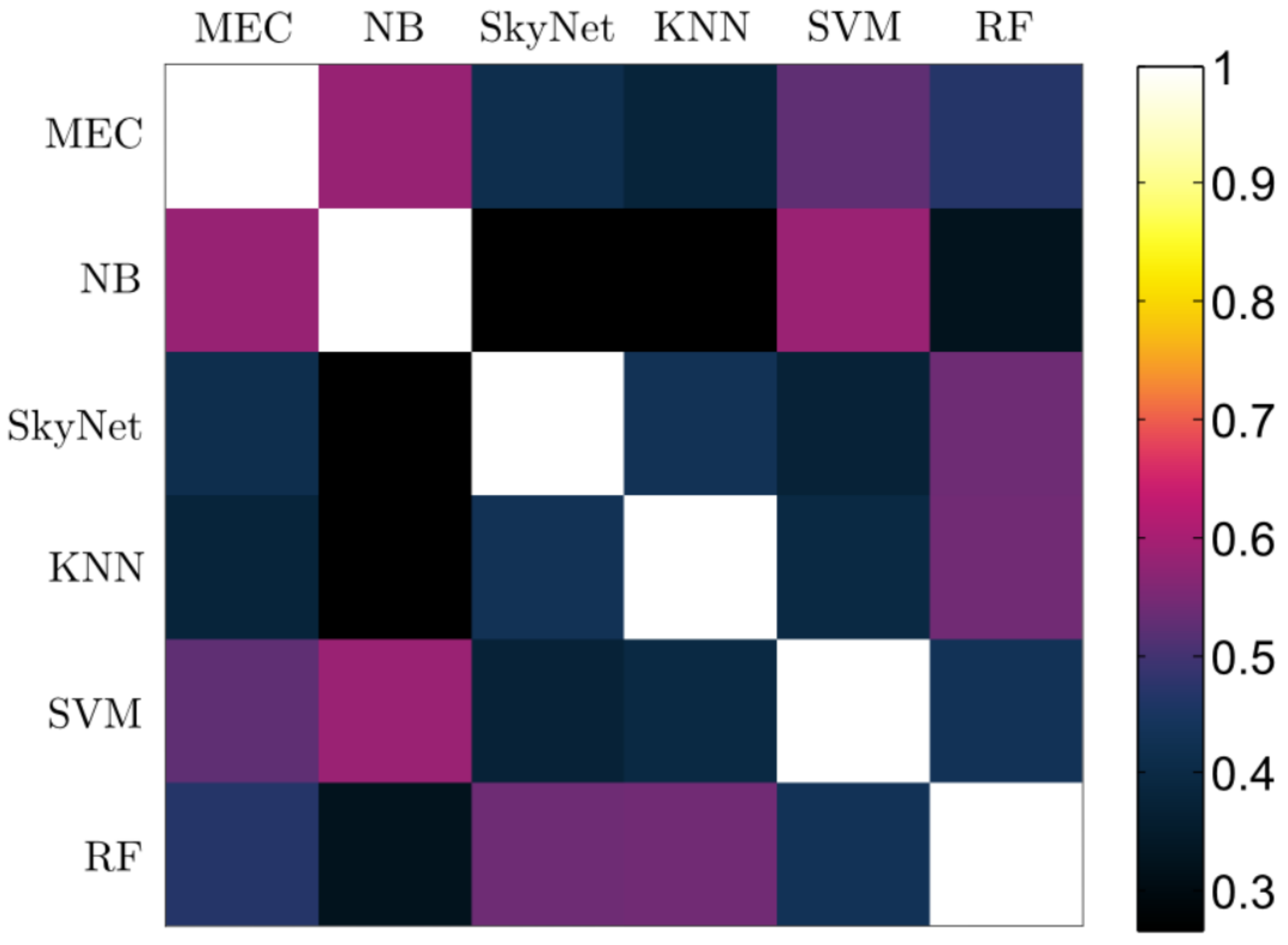}
\caption{The Cohen Kappa coefficient ($\kappa$) value for each pair of classifiers, which measures the overlap in performance between classifiers relative to pure chance (which corresponds to $\kappa = 0$ ). These results can be interpreted with the aid of Table \ref{tab:interp}.}
\label{fig:kappa}
\end{figure}

Fig. \ref{fig:kappa} shows the $\kappa$-values calculated for all pairs of classifiers. It can be seen that the strongest agreement between any two classifiers is "moderate" (from Table \ref{tab:interp}), and is found between SVM and NB, and MEC and NB. The fact that none of the top algorithms show strong agreement suggests that it might be fruitful to combine the predictions of different classifiers together with an ensemble classifier.

\subsection{Classifier Performance on Different Classes of Images}
\label{class_performance}
The performance of the classifiers on the various different visual classes (real objects, artefacts and dipoles/saturated) in the best performing test data sets is shown in Table \ref{tab:class} (note that recall, here, has the same meaning as accuracy). For the calculation of the recall values, it should be noted that true positives ($t_p$) here correspond to the objects in a certain class classified correctly as either real or not-real, while false negatives ($f_n$) correspond to incorrect classifications.

For all classifiers the performance on real objects and artefacts is higher than that on the dipoles/saturated class, the latter therefore being the class that lowers the overall performance of the classifiers. The relatively poor performance on this class can be understood by noting that, firstly, the objects have features that can be quite similar to those of real objects (this can be seen when comparing the PCs in Fig. \ref{fig:pca_satdip} to those in Fig. \ref{fig:pca_real}), and will therefore confuse classifiers. On further visual inspection it can be noted that the images in this class (in general) shows a greater diversity in features than real object images do, leading to there not being as strong a common feature to characterise almost all of the images with, as is the case with real object images. When an object in the dipole/saturated class is encountered that looks similar to a real object (this is often the case for noisy images), it can therefore easily be misclassified as strong features of the real object images may approximate the dipole/saturated image more closely than weaker/mixed features from the dipole/saturated class itself. Section \ref{incorrect} further discusses misclassification issues. 

\begin{table}
\centering
 \caption{The recall of the various classifiers on the different classes of images. Here the true positives, $t_p$, for a class corresponds to the images in that class that were correctly classified, and $f_n$ corresponds to the images that were incorrectly classified. The best results are indicated in bold.}
 \label{tab:class}
 {\renewcommand{\arraystretch}{1.5}
 \begin{tabular}{@{}lcccccc}
  \hline
  Class & RF & KNN & SkyNet & SVM & MEC & NB \\
  \hline
  \hline
  Real & 0.91 & 0.90 & 0.89 & 0.90 & {\bf 0.92} & 0.86 \\
  Artefact & {\bf 0.97} & 0.93 & 0.92 & 0.90 & 0.90 & 0.82 \\
  Dip/Sat & {\bf 0.80} & 0.79 & 0.76 & 0.62 & 0.47 & 0.38 \\
  \hline
 \end{tabular}}
\end{table}

\subsection{Examples of Incorrectly Classified Objects}
\label{incorrect}
\begin{figure}
\centering
\begin{subfigure}{.16\textwidth}
  \centering
  \includegraphics[width=\linewidth]{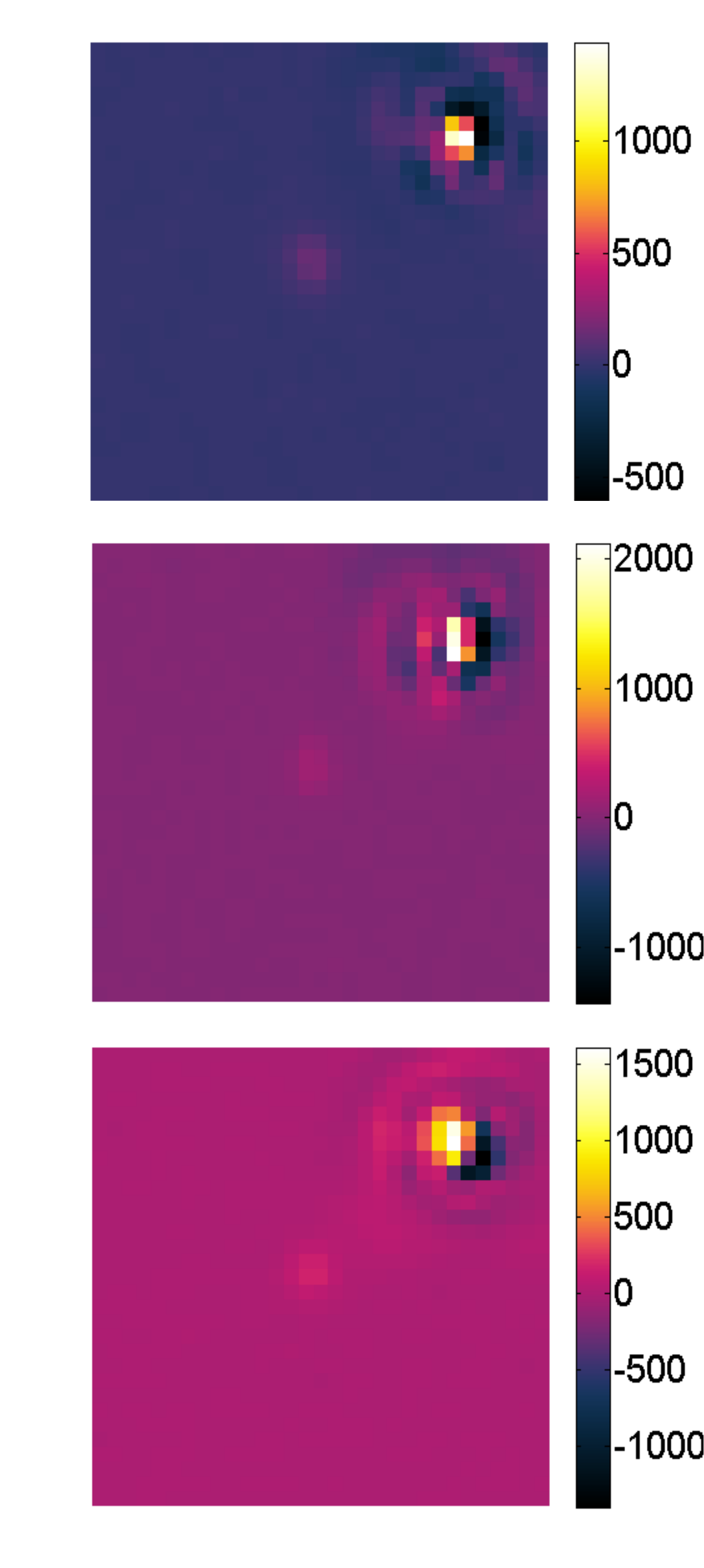}
  \caption{}
  \label{fig:real1}
\end{subfigure}%
\begin{subfigure}{.16\textwidth}
  \centering
  \includegraphics[width=\linewidth]{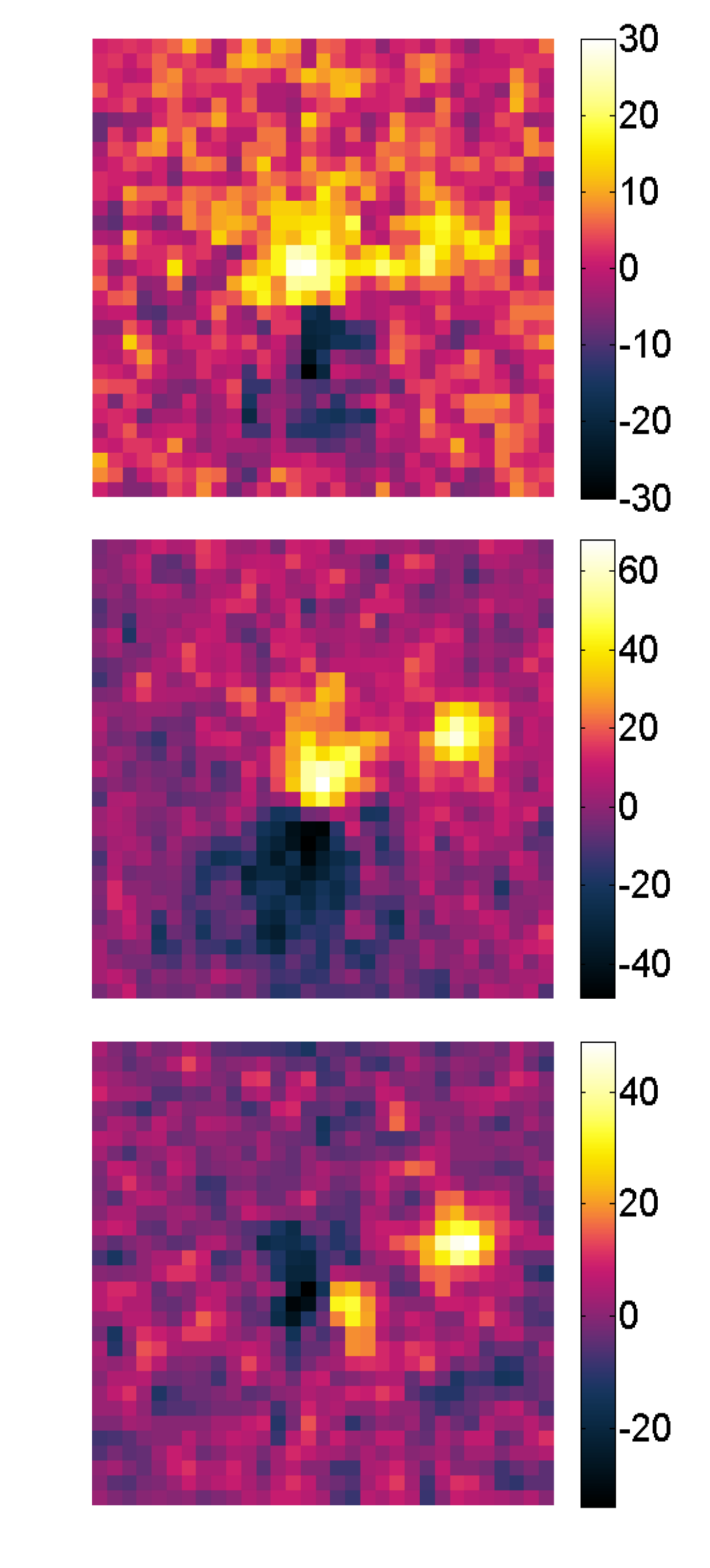}
  \caption{}
  \label{fig:real2}
\end{subfigure}%
\begin{subfigure}{.16\textwidth}
  \centering
  \includegraphics[width=\linewidth]{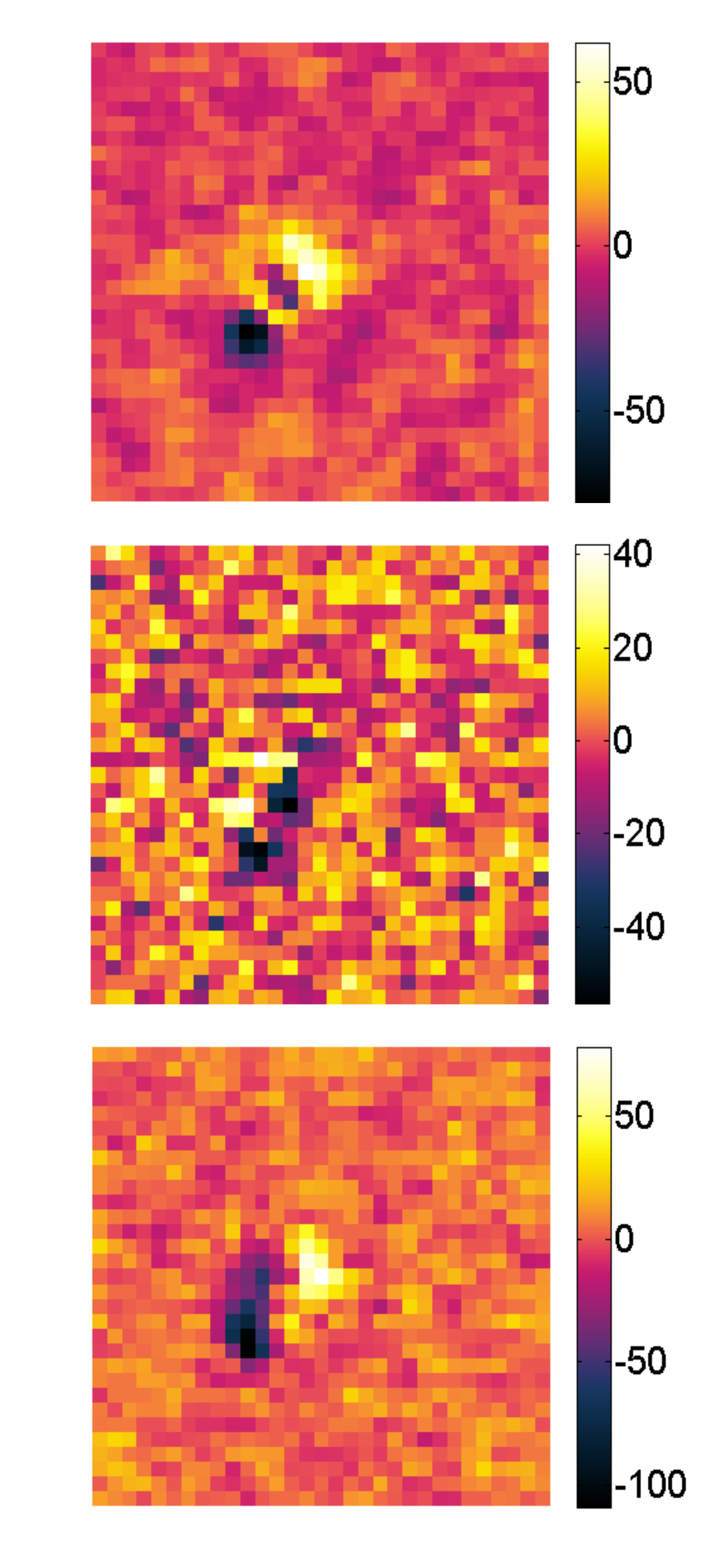}
  \caption{}
  \label{fig:real3}
\end{subfigure}%
\caption{Examples of objects classified as real by the human scanners but classified as not-real by all six machine classifiers; for each image the {\it g}-band (top), {\it i}-band (middle) and {\it r}-band (bottom) image is shown: (a) the dipole-like residuals in the top corner outshining the faint point-like structures probably resulted in misclassification; (b\&c) the dipole-like structures observed in the centre of these real object images probably resulted in their misclassification.}
\label{fig:real_incorrect}
\end{figure}

\begin{figure}
\centering
\begin{subfigure}{.16\textwidth}
  \centering
  \includegraphics[width=\linewidth]{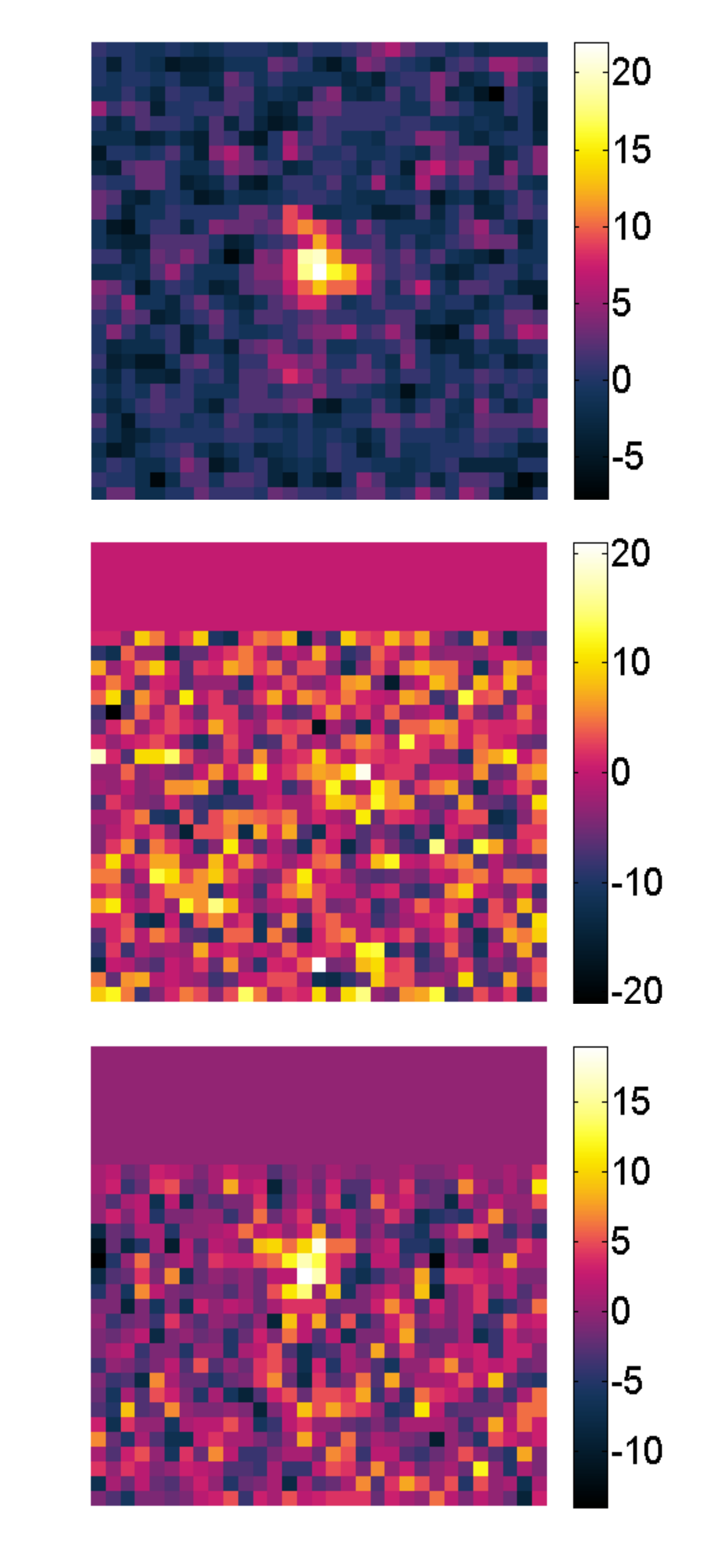}
  \caption{}
  \label{fig:notreal1}
\end{subfigure}%
\begin{subfigure}{.16\textwidth}
  \centering
  \includegraphics[width=\linewidth]{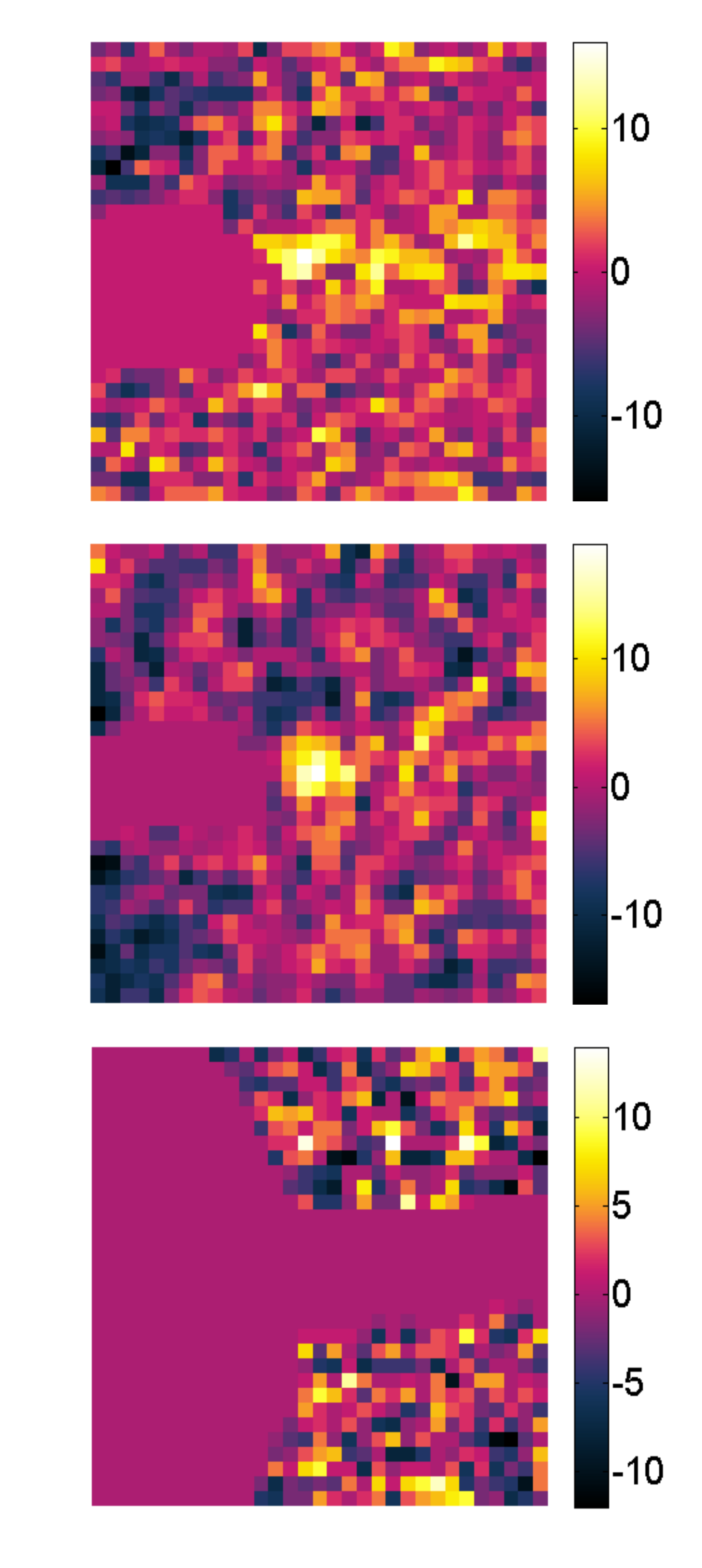}
  \caption{}
  \label{fig:notreal2}
\end{subfigure}%
\begin{subfigure}{.16\textwidth}
  \centering
  \includegraphics[width=\linewidth]{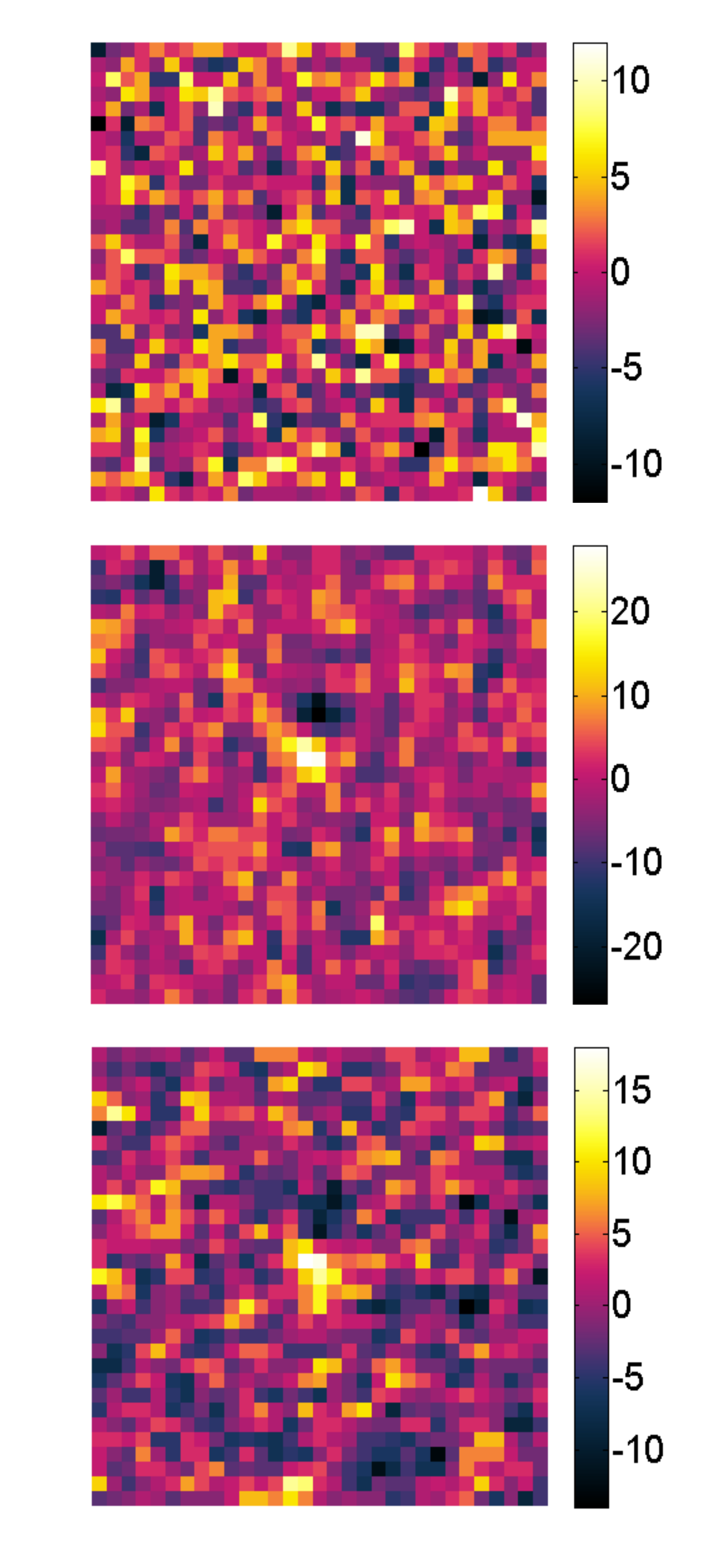}
  \caption{}
  \label{fig:notreal3}
\end{subfigure}%
\caption{Examples of objects classified as not-real by humans that were each classified as being real by all six machine classifiers; for each \mbox{image} the {\it g}-band (top), {\it i}-band (middle) and {\it r}-band (bottom) image is shown: (a\&b) these artefacts were probably misclassified due to the fact that they appear more point-like than spike-like; (c) the noisy images of this dipole probably resulted in its misclassification.}
\label{fig:notreal_incorrect}
\end{figure}

To give an idea of what causes certain images to get misclassified, we inspected images that were each incorrectly classified by all six classifiers. Fig. \ref{fig:real_incorrect} shows examples of real objects that were classified as being not-real and Fig. \ref{fig:notreal_incorrect} shows examples of not-real objects that were classified as being real.

In Fig. (\ref{fig:real1}) it is observed that all three passbands have a faint point-like structure at the centre of their images; this is the reason for the object's 'real' label, as these point-like residuals point towards there being a possible SN. The reason for its misclassification as a not-real object is probably due to the much brighter dipole-like residual in the top right corner of all the images, something that would confuse the classifiers since we did not force them to focus only on the central few pixels. Fig. (\ref{fig:real2}) and (\ref{fig:real3}) both show dipole-like structures in all three bands instead of the usual point-like residuals characteristic of real objects - it is no wonder then that all six classifiers misclassified these objects as being not-real. The original classification of these two objects as real might even be a mistake on the part of the hand scanners, which emphasises how errors in the labelling of the training data can propagate through to the test set. 

The same sort of problem is evident in Fig. (\ref{fig:notreal1}) where human scanners classified an object as an artefact that the classifiers understandably mistook for a real object due to the point-like residuals observed in all bands. Fig. (\ref{fig:notreal2}) is an artefact with masked areas in all three bands causing the object to lose most of its spike-like attributes, something very characteristic of the artefact subclass. The fact that all six classifiers classified this object as real is possibly mostly due to the point-like structure at the centre of the {\it i}-band image. Fig. (\ref{fig:notreal3}) is a dipole, as can be seen by the presence of dipole-like residuals in the {\it i} and {\it r}-bands. The fact that these images are quite noisy (especially the {\it g}-band image) possibly contributed to the fact that all classifiers predicted it to be a real object. Furthermore, the features of the real and dipole/saturated objects are quite similar in some cases, a possible further contribution to the misclassification of the object as real instead of not-real (this problem was discussed in section \ref{class_performance}).

\

\

\subsection{Performance on Spectroscopically Confirmed SNe}
The performance of the classifiers on the spectroscopically confirmed SNe in the best performing test data feature sets is shown in Table \ref{tab:spec}. Out of the 135 spectroscopically confirmed SNe found in our testing data, 110 were SNe Ia and 19 were other SNe (type Ib, Ic and II).

It is clear that RF, KNN, SkyNet, SVM and MEC performs well, with the lowest recall value amongst them at 0.90 for SNe Ia, echoing earlier successes. NB continues to be the poorest performer of the group, with its highest recall value at 0.83 for spectroscopically confirmed SNe Ia.

\begin{table}
\centering
 \caption{The recall of the various classifiers on the 135 spectroscopically confirmed SNe in the test set (110 SNe Ia and 19 other SNe). The first row shows the performance on the group of SNe Ia, while the second row details the results for the other 19 SNe (excluding SNe Ia). The last row shows the classification results for all 135 spectroscopically confirmed SNe together. The best results are shown in bold.}
 \label{tab:spec}
 {\renewcommand{\arraystretch}{1.5}
 \begin{tabular}{@{}lcccccc}
  \hline
    & RF & KNN & SkyNet & SVM & MEC & NB \\
  \hline
  \hline
  SN Ia & 0.90 & {\bf 0.94} & 0.92 & 0.90 & 0.91 & 0.83 \\
  SNe & {\bf 1.00} & {\bf 1.00} & {\bf 1.00} & 0.89 & {\bf 1.00} & 0.63 \\
  All & 0.91 & {\bf 0.95} & 0.93 & 0.90 & 0.93 & 0.80 \\
  \hline
 \end{tabular}}
\end{table}

\subsection{Comparison with Human Scanners}
As part of the SDSS SN survey fake SNe were inserted into the pipeline to test the efficiency of the hand scanners. The fake SN-tag recall, averaged over scanners, was $0.956 \pm 0.010$ \citep{pvt1}. In comparison, Table \ref{tab:fakes} shows the classification performance of the various classifiers on the same fake SNe found in the test set. It can be seen that all the algorithms other than NB perform comparably to, or even better than, the average human scanner on the fake SNe, with RF leading with a recall of 0.97. Furthermore, when looking at Fig. \ref{fig:roc}, it can be seen that by changing the discrimination threshold value and inducing only slight penalties (in the form of a larger FPR), a recall (TPR) value of 96\% can easily be achieved even when all the test data is used, outmatching the human classifiers.

\begin{table}
\centering
 \caption{A summary of the performance results of the various classifiers on the fake SNe found in the test set. The best result is indicated in bold.}
 \label{tab:fakes}
 {\renewcommand{\arraystretch}{1.5}
 \begin{tabular}{@{}lcccccccc}
  \hline
  Machine Learning Technique & Recall\\
  \hline
  \hline
  RF & {\bf 0.97}\\
  KNN & 0.96\\
  SkyNet & 0.96\\
  SVM & 0.94\\
  MEC & 0.96\\
  NB & 0.90\\
  \hline
 \end{tabular}}
\end{table}

\section{Conclusions and Future Work}
In examining a broad spectrum of machine learning techniques for transient classification we have not tried not only to find an optimal solution for the problem but also to demonstrate that it is relatively easy to match or exceed human efficiency with non-linear classifiers such as random forests, artificial neural networks, support vector machines and k-nearest neighbours. The impressive success of even the linear minimum error classifier suggests that much of the heavy lifting here is done by the judicious use of (class-based) PCA as the basis for feature extraction. Overall we found that random forests performed best in all metrics other than recall on the full test set and also performed best on the fake SNe, in line with their superior performance in \cite{p24}. Their performance on the small spectroscopically-confirmed subset of SN Ia, however, lagged behind KNN, MEC and the artificial neural networks.

In addition to the obvious advantages that machine learning techniques offer in terms of handling large volumes of data, we also stress that automated classifiers have quantified and controllable errors and biases. This latter feature will be essential for building pipelines for data analysis which fully propagate all systematic errors in a reproducible way.

Having illustrated that PCA feature extraction plus a simple classifier provides a robust transient identification solution, the next step would be to build this approach into the data analysis for current and next-generation experiments. This will involve tailoring and optimizing the classification pipeline for further performance enhancements. One option would be to combine classifiers (through e.g. boosting or a weighted voting system); given the fact that the classifiers do not agree that strongly (see Fig. \ref{fig:kappa}) this could be a fruitful avenue to explore. Another strategy would be to try and improve the efficiency for identifying dipole/saturated images (consistently the worst classified category) by augmenting the feature set.

Finally we note that since we have only used difference images we have not used any host galaxy information, nor have we used the relative colours of the bands nor multi-epoch data. All of these can be studied and will prove particularly useful for general-purpose surveys that look to classify objects beyond just two or three classes.

\section*{Acknowledgements}
BB and MS thank our collaborators in the SDSS supernova survey team for insights gained during the duration of the survey. LdB thanks Johan A. du Preez and Ben Herbst for their insights during the initial phases of the project. The authors further thank the University College London (UCL) Physics and Astronomy Department for the use of their Splinter computer cluster. LdB is supported by a South African Square Kilometre Array (SKA) Project postgraduate scholarship; NS is supported by a Claude Leon Foundation fellowship; BB acknowledges funding from the South African National Research Foundation (NRF) and SKA; MS is supported by the South African SKA Project, the NRF and the UK Science \& Technology Facilities Council (STFC).

Funding for the SDSS and SDSS-II has been provided by the Alfred P. Sloan Foundation, the Participating Institutions, the National Science Foundation, the U.S. Department of Energy, the National Aeronautics and Space Administration, the Japanese Monbukagakusho, the Max Planck Society, and the Higher Education Funding Council for England. The SDSS Web Site is http://www.sdss.org/.

The SDSS is managed by the Astrophysical Research Consortium for the Participating Institutions. The Participating Institutions are the American Museum of Natural History, Astrophysical Institute Potsdam, University of Basel, Cambridge University, Case Western Reserve University, University of Chicago, Drexel University, Fermilab, the Institute for Advanced Study, the Japan Participation Group, Johns Hopkins University, the Joint Institute for Nuclear Astrophysics, the Kavli Institute for Particle Astrophysics and Cosmology, the Korean Scientist Group, the Chinese Academy of Sciences (LAMOST), Los Alamos National Laboratory, the Max-Planck-Institute for Astronomy (MPIA), the Max- Planck-Institute for Astrophysics (MPA), New Mexico State University, Ohio State University, University of Pittsburgh, University of Portsmouth, Princeton University, the United States Naval Observatory, and the University of Washington.

\bibliographystyle{mn2e}
\bibliography{main}

\section*{Appendix}
This section assists in describing the classification system used by the SDSS hand-scanners during visual inspection of images (see section \ref{data}). While Table \ref{tab:class_cat} describes each of the different classes, Table \ref{tab:hand_class} shows how the objects in each of these classes look like. It shows six {\it r}-band difference images from each of the ten classes and also illustrate the difference in image-quality based on the signal-to-noise ratio.

\begin{table*}
\centering
 \caption{The ten classes, described in Table \ref{tab:class_cat}, used by the SDSS hand-scanners during visual classification. For each class, six {\it r}-band difference images of objects are shown - three of which have an SNR above 40, and three having an SNR below 20 (representing roughly 80 \% of the dataset). Images were selected in such a way that they provide a faithful representation of their respective classes.}
 \label{tab:hand_class}
 {\renewcommand{\arraystretch}{2}
 \begin{tabular}{N|M|M} 
  \hline
  Original Class & SNR $> 40$ & SNR $< 20$ \\
  \hline
  \hline
  {\bf Artefact} & \includegraphics[scale=0.35]{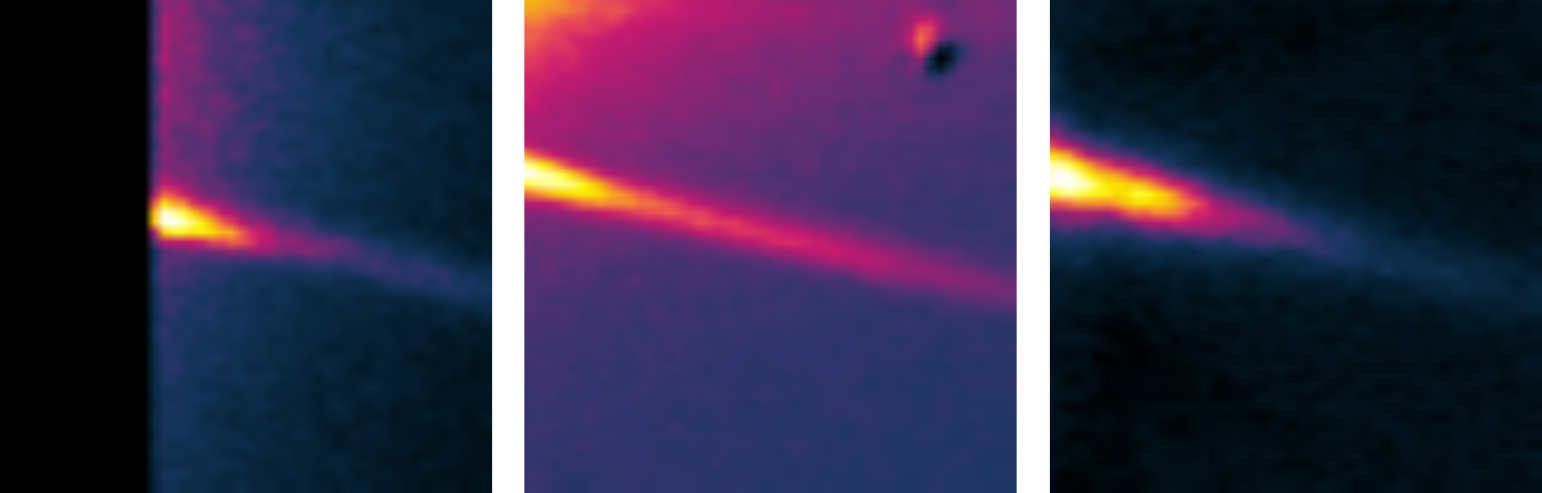} & \includegraphics[scale=0.35]{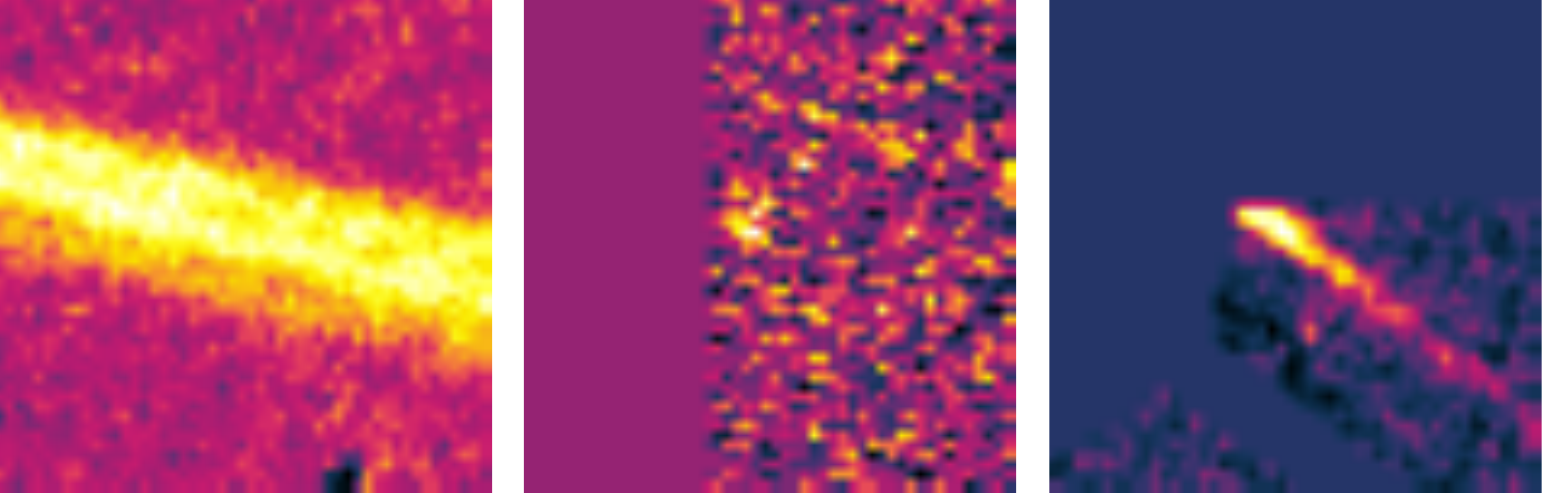} \\
  
  {\bf Dipole} & \includegraphics[scale=0.35]{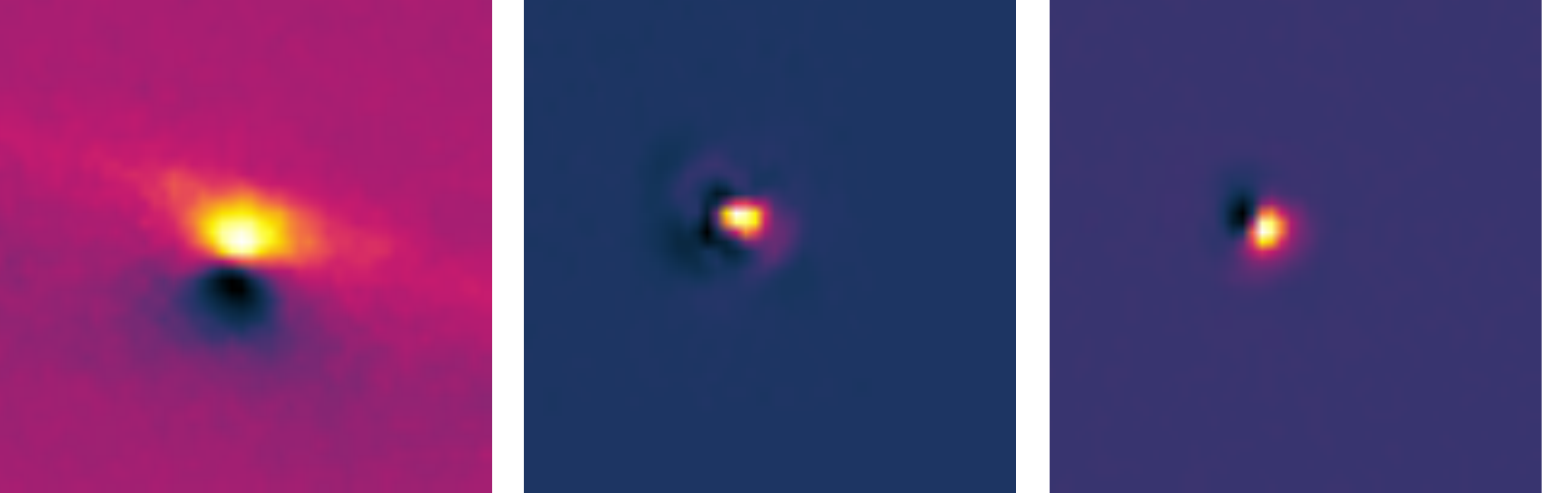} & \includegraphics[scale=0.35]{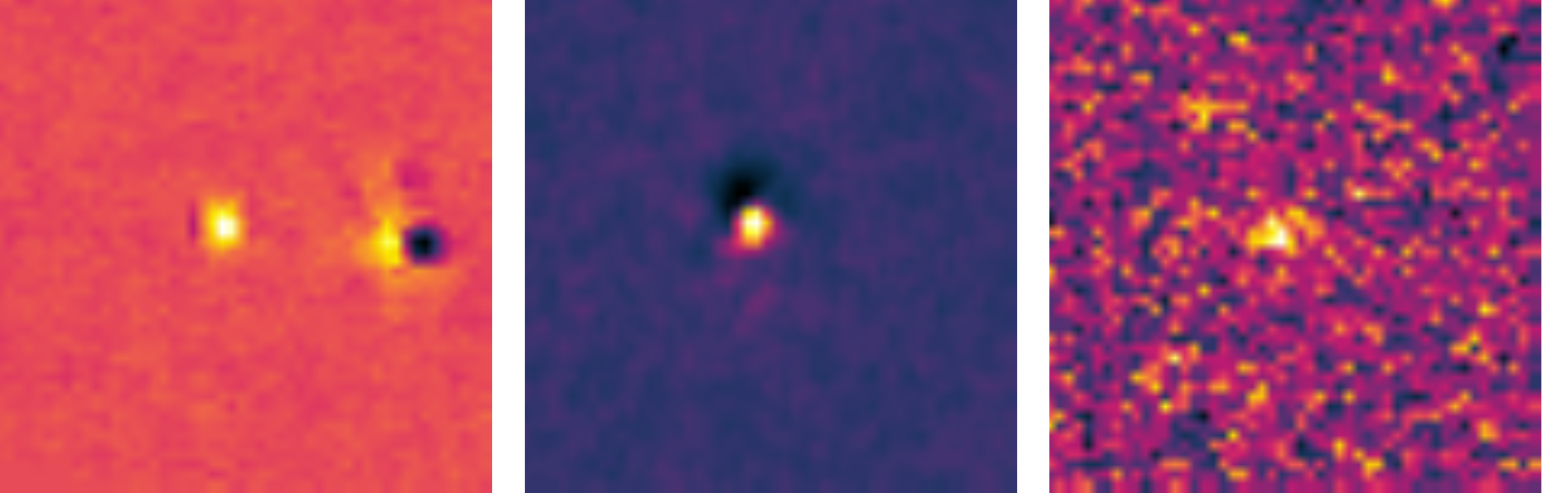} \\
  
  {\bf Saturated Star} & \includegraphics[scale=0.35]{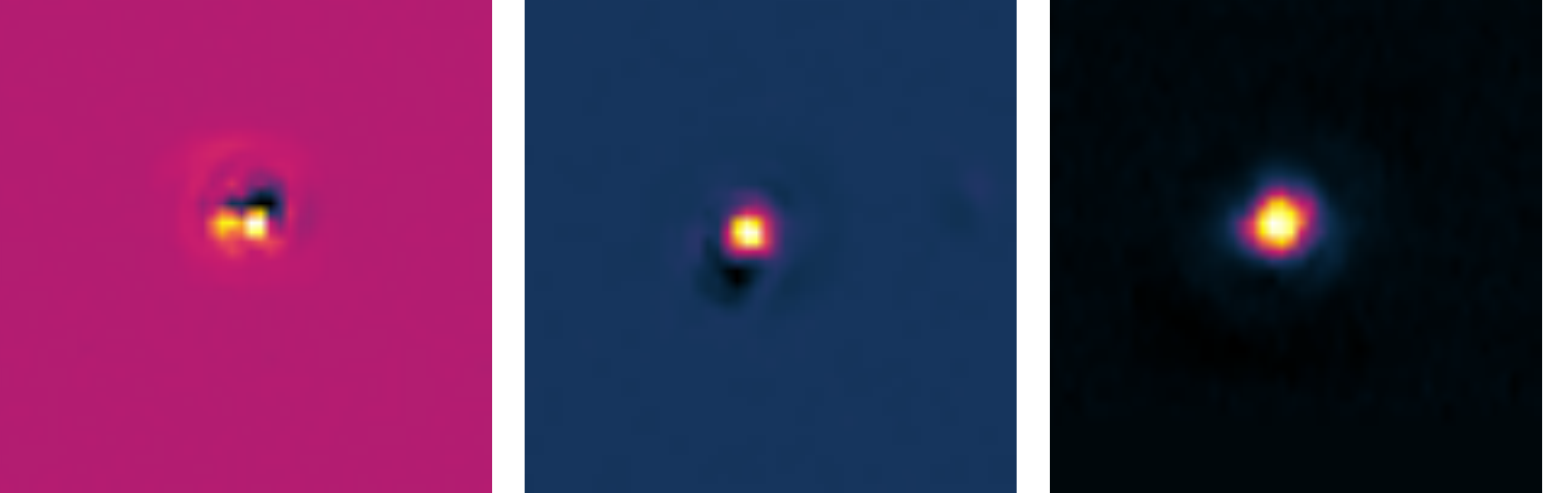} & \includegraphics[scale=0.35]{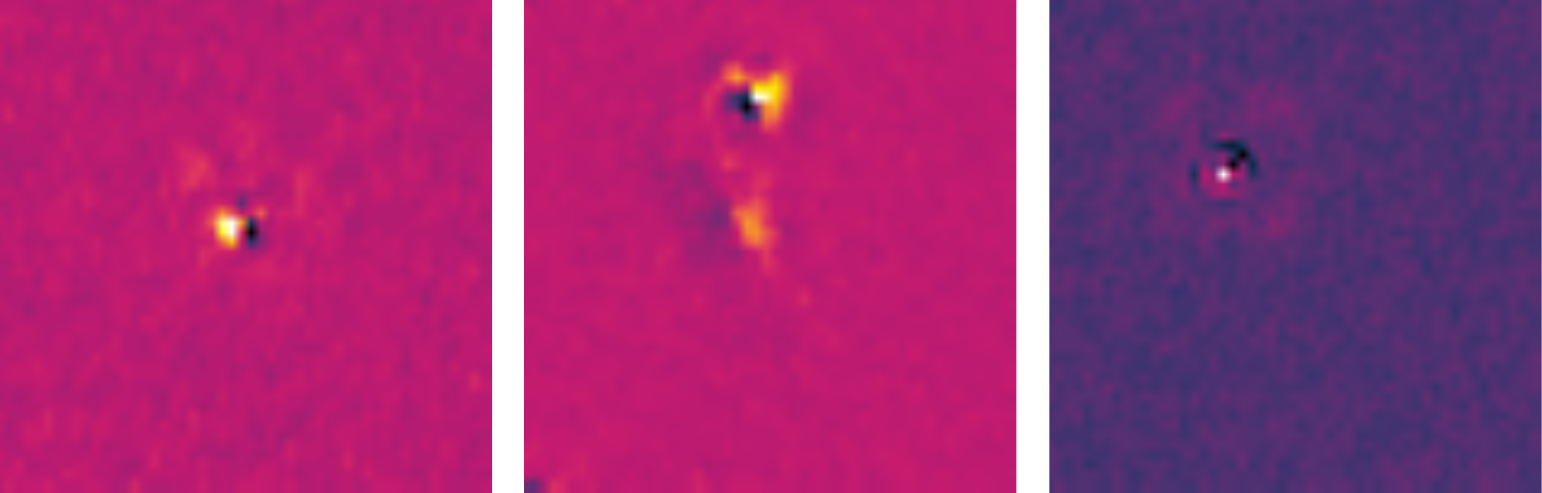} \\
  
  {\bf Moving} & \includegraphics[scale=0.35]{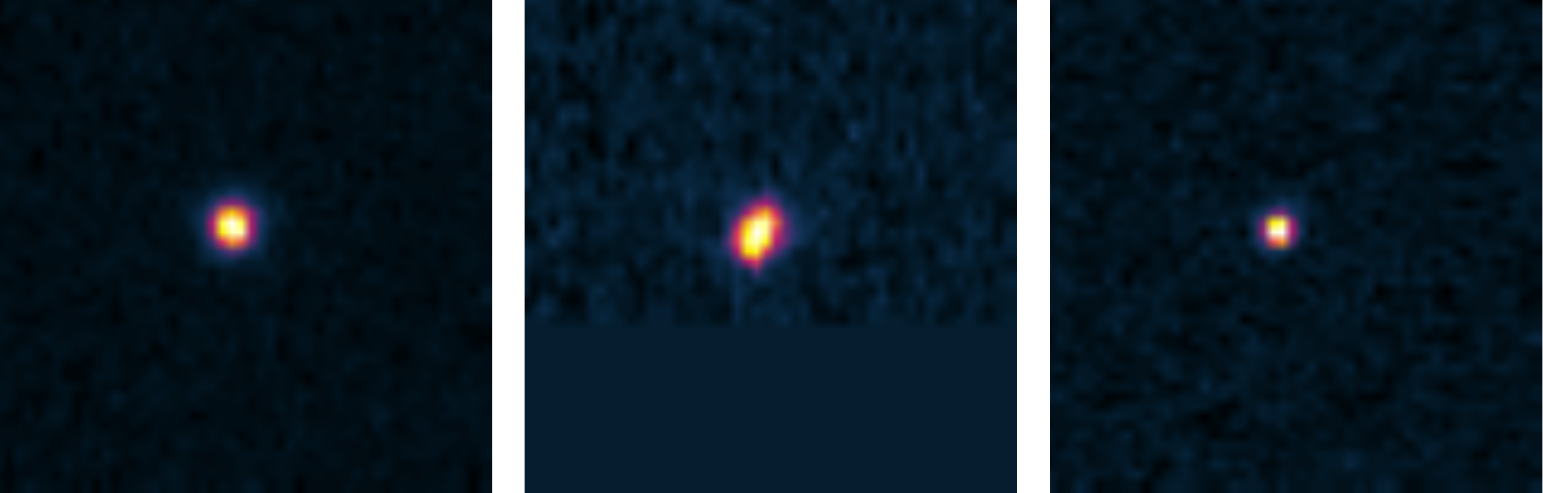} & \includegraphics[scale=0.35]{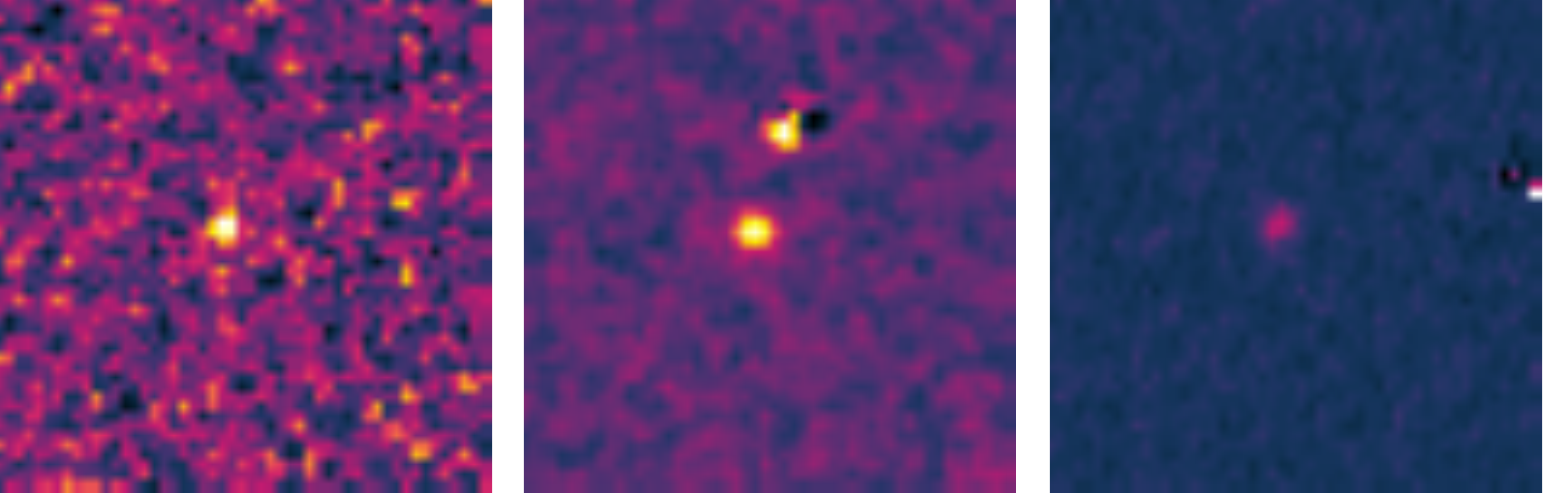} \\
  
  {\bf Variable} & \includegraphics[scale=0.35]{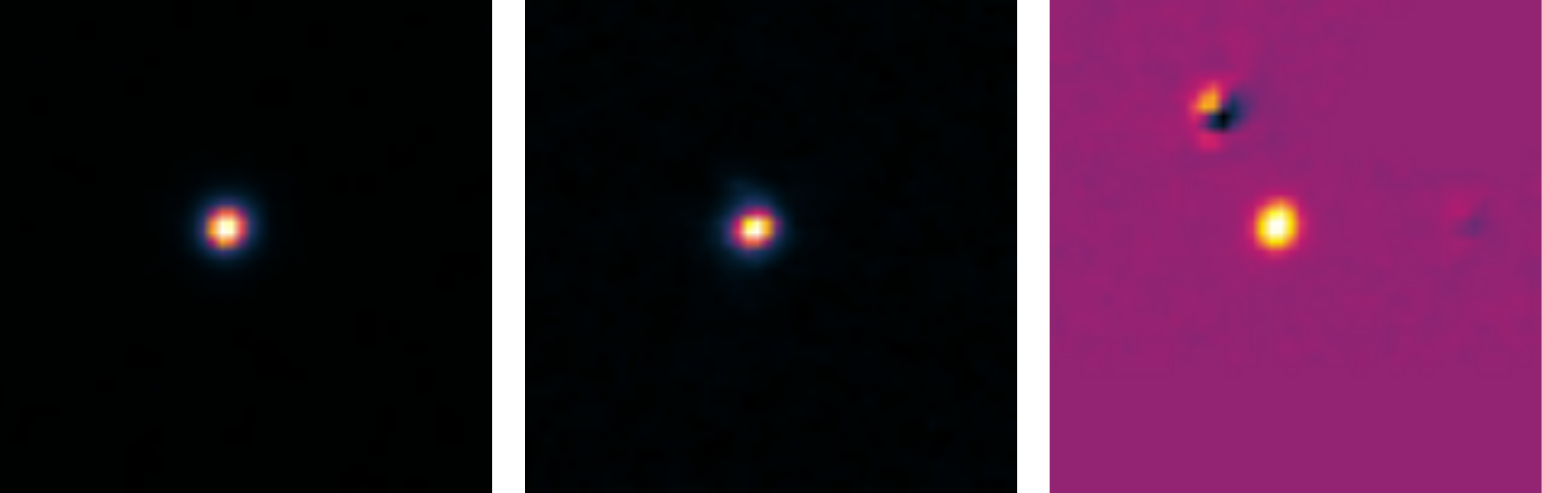} & \includegraphics[scale=0.35]{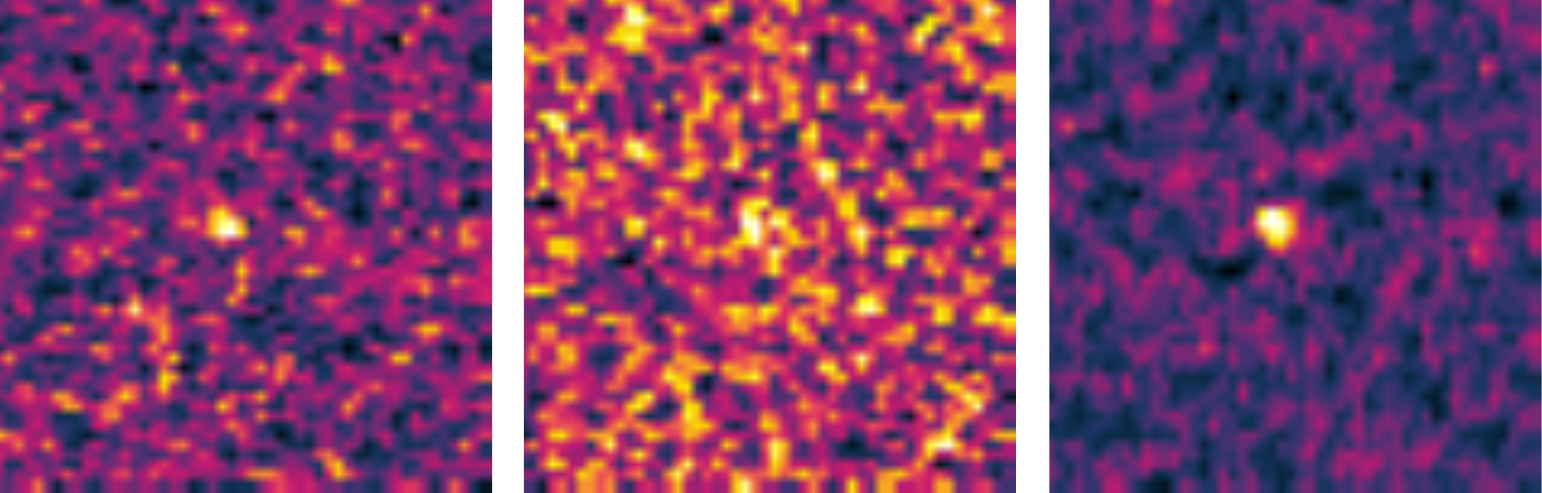} \\
  
  {\bf Transient} & \includegraphics[scale=0.35]{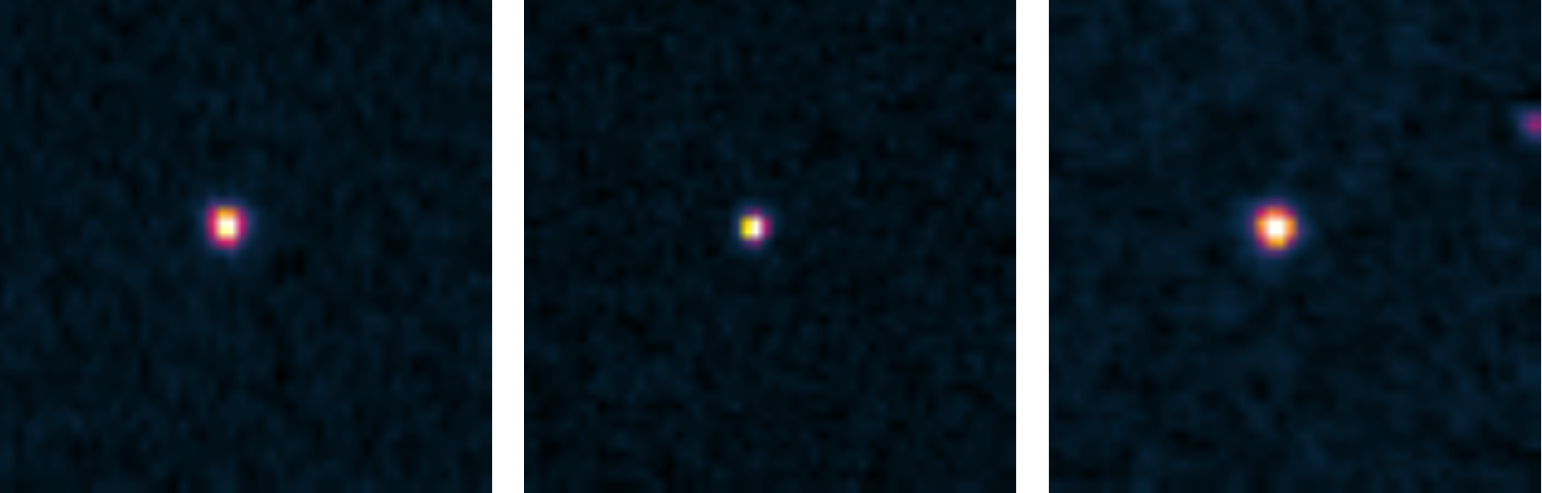} & \includegraphics[scale=0.35]{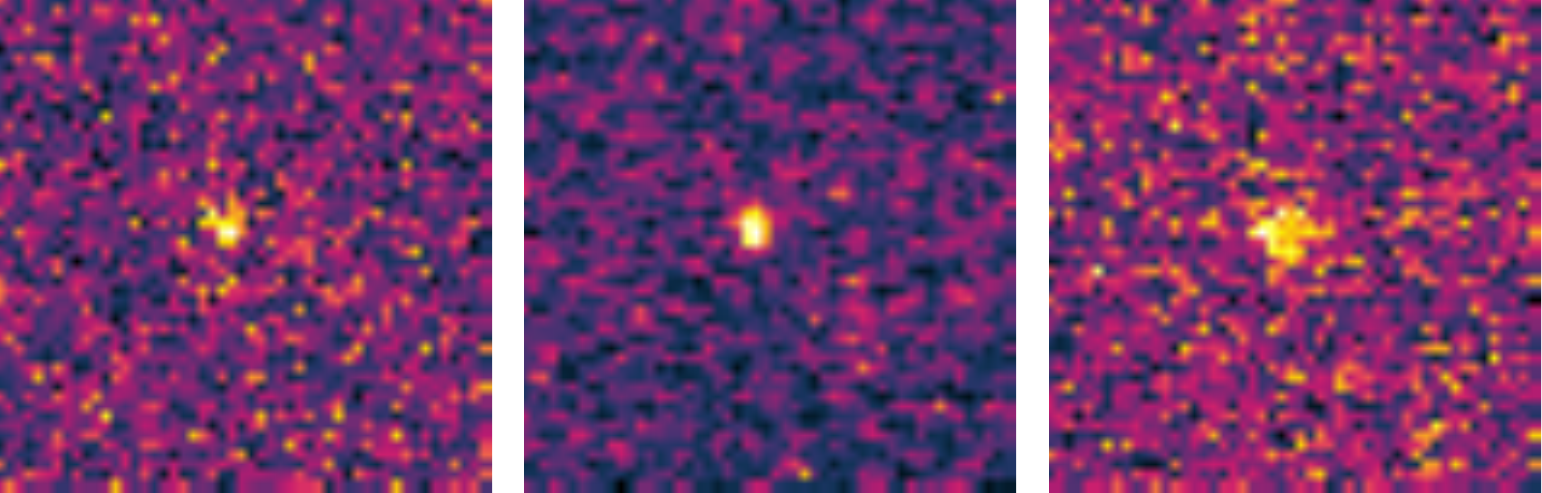} \\
  
  {\bf SN Other} & \includegraphics[scale=0.35]{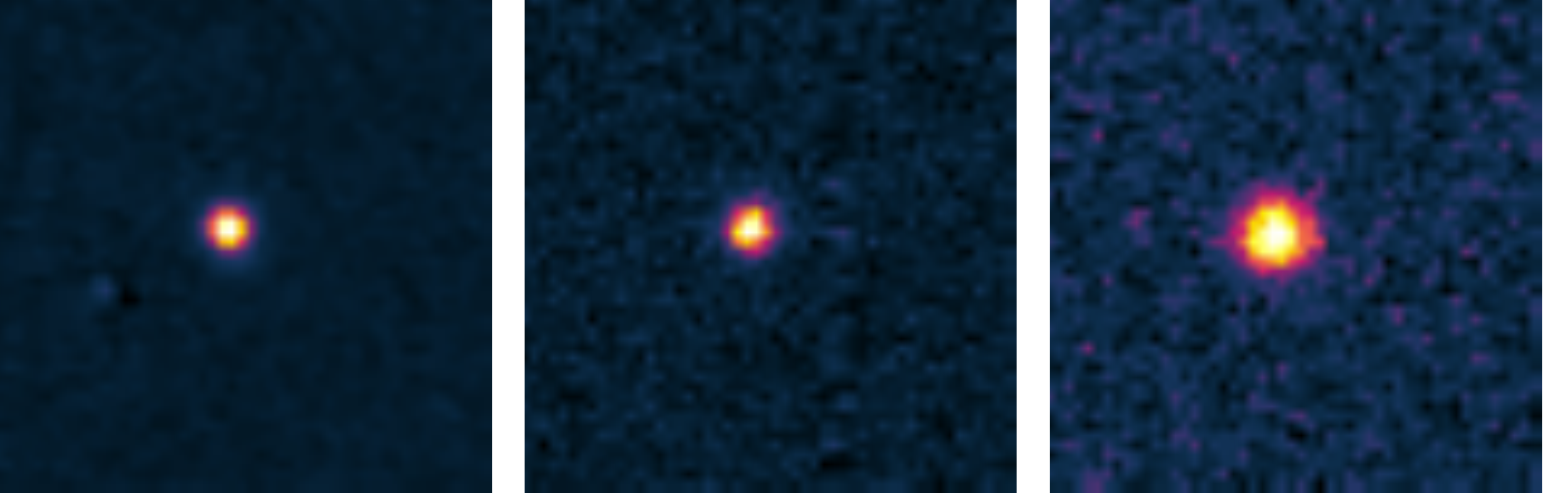} & \includegraphics[scale=0.35]{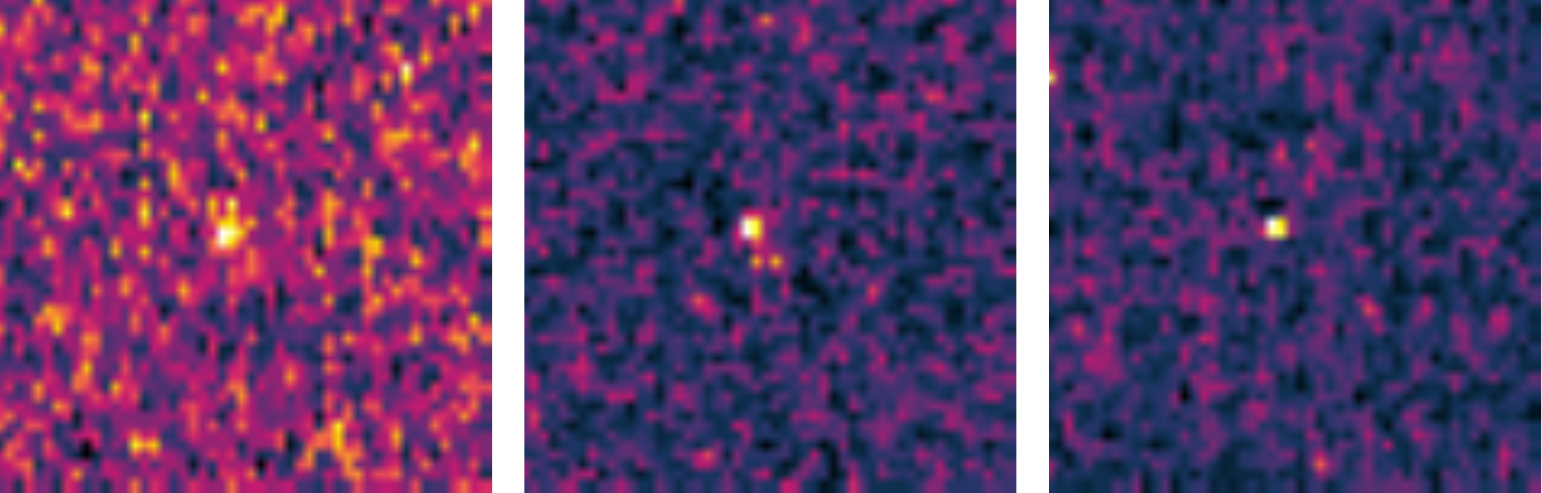} \\
  
  {\bf SN Bronze} & \includegraphics[scale=0.35]{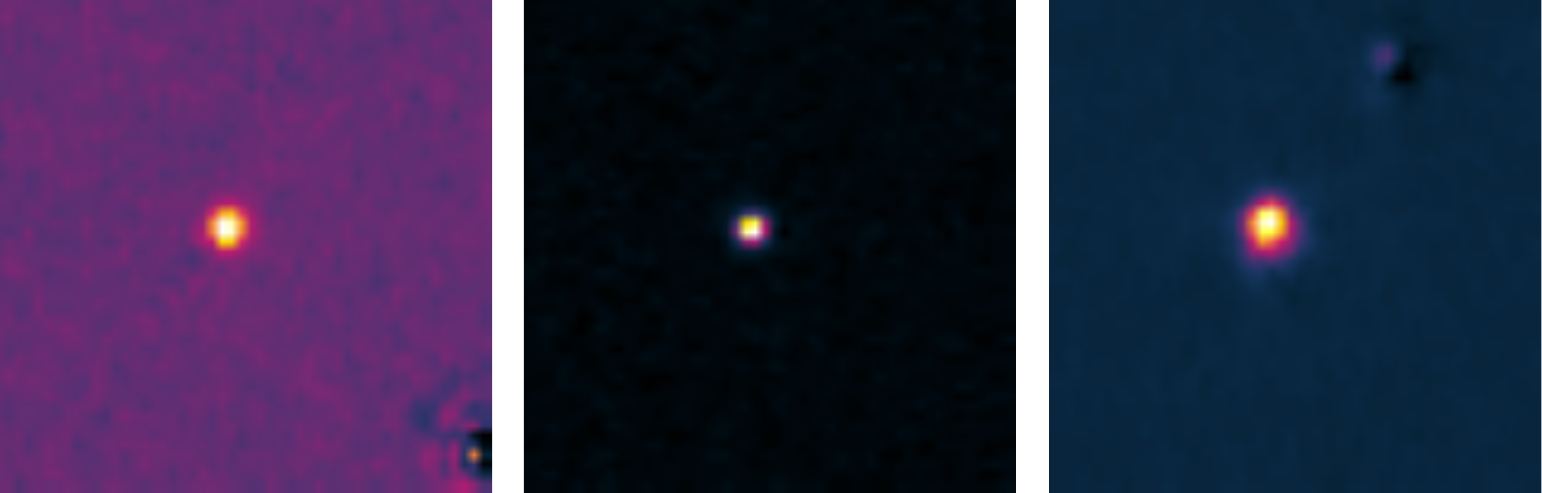} & \includegraphics[scale=0.35]{bro_good.pdf} \\
  
  {\bf SN Silver} & \includegraphics[scale=0.35]{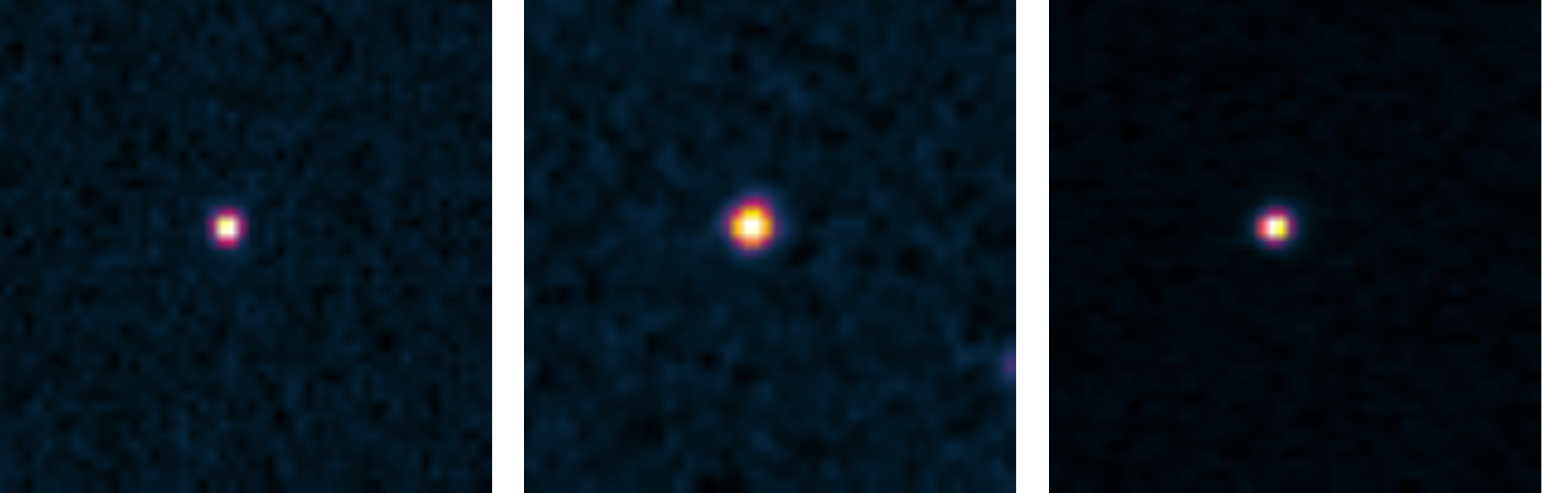} & \includegraphics[scale=0.35]{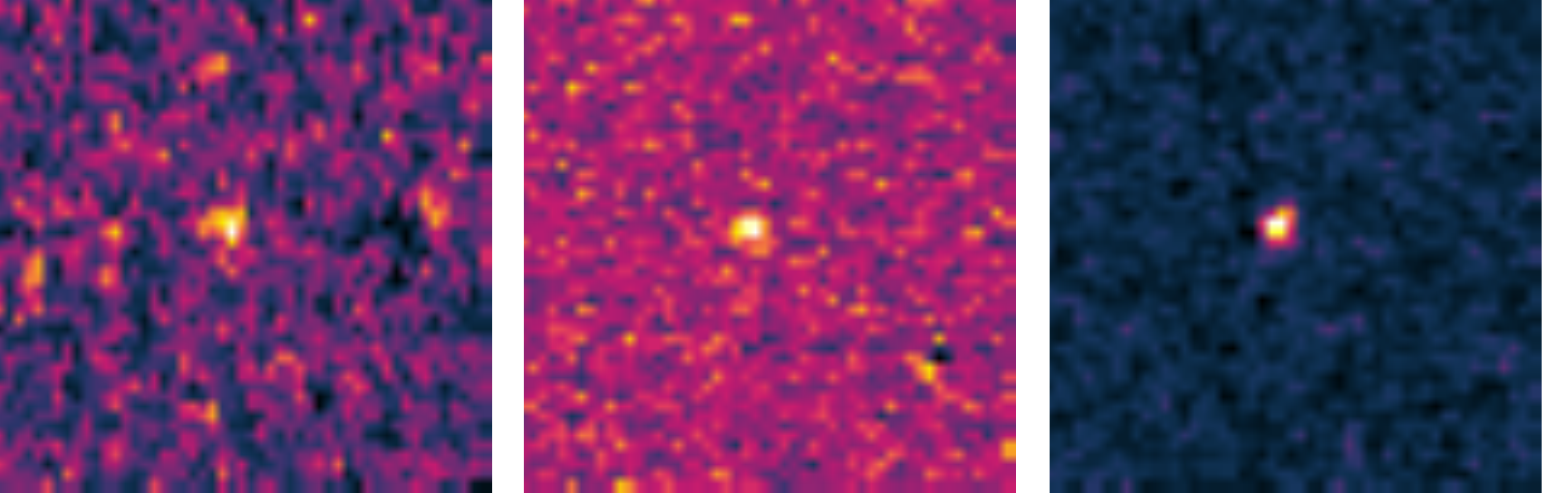} \\
  
  {\bf SN Gold} & \includegraphics[scale=0.35]{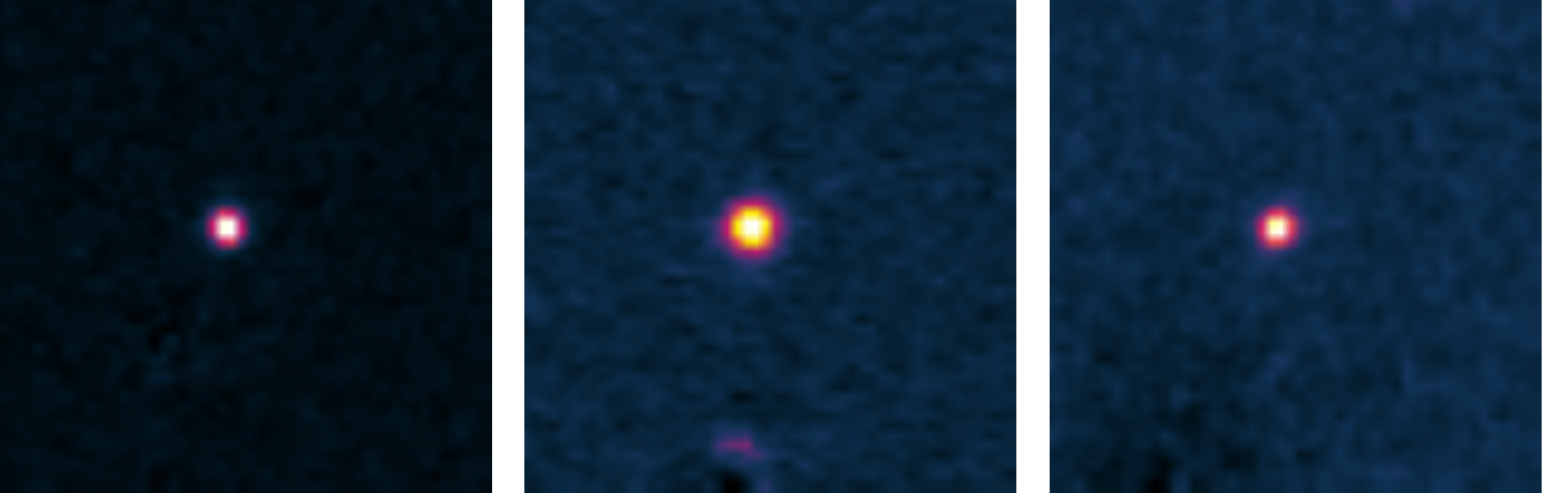} & \includegraphics[scale=0.35]{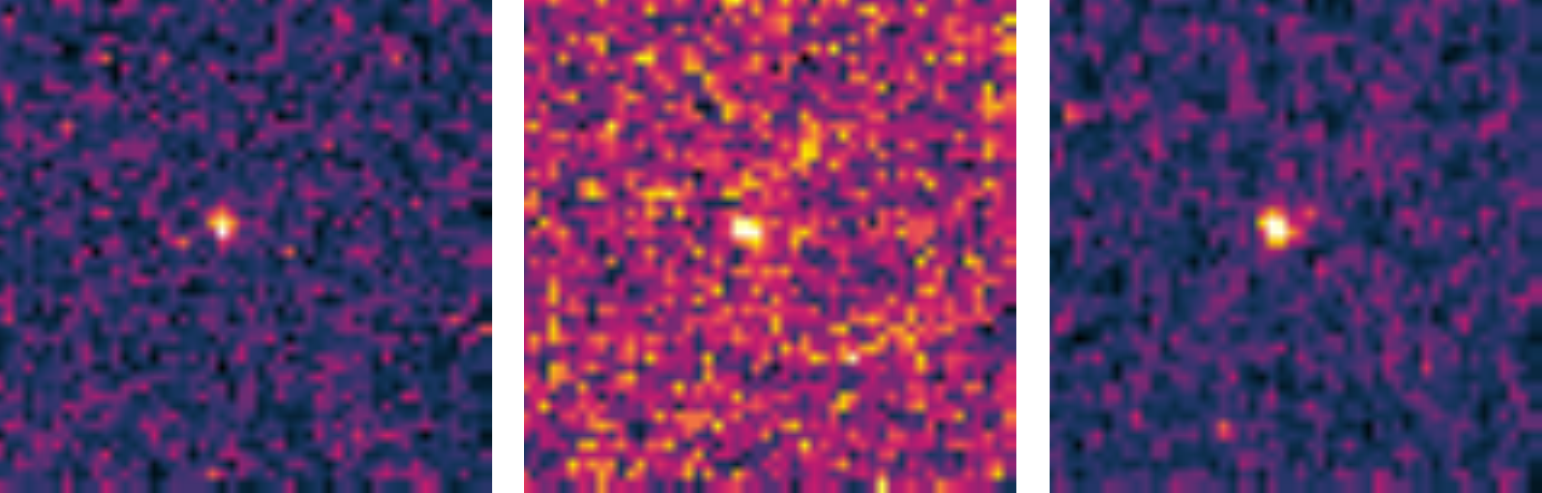} \\
  \hline
 \end{tabular}}
\end{table*}

\end{document}